\documentclass[12pt]{article}
\pdfoutput=1
\usepackage{amsmath,amssymb,amsthm,amsxtra,overpic,bbm,bm,epsfig,subfigure}
\usepackage{hyperref}
\usepackage{graphicx}
\usepackage{color}
\usepackage{array}
\usepackage{comment}
\usepackage{epstopdf}
\usepackage{float}
\usepackage{multirow}
\numberwithin{equation}{section}
\usepackage{cite}
\usepackage{hyperref}
\usepackage{slashed,stmaryrd}
\usepackage{bbm}
\usepackage{lscape}%
\usepackage{array}
\usepackage{booktabs}%

\def\thefootnote{\fnsymbol{footnote}}

\begin{document}

\vspace{0.2cm}

\begin{center}
{\Large\bf Explicit Perturbations to the Stabilizer $\tau = {\rm i}$ of Modular $A^\prime_5$ Symmetry and Leptonic CP Violation}
\end{center}

\vspace{0.2cm}

\begin{center}
	{\bf Xin Wang}~$^{a,~b}$~\footnote{E-mail: wangx@ihep.ac.cn},
	\quad
	{\bf Shun Zhou}~$^{a,~b}$~\footnote{E-mail: zhoush@ihep.ac.cn (corresponding author)}
	\\
	\vspace{0.2cm}
	{\small
		$^a$Institute of High Energy Physics, Chinese Academy of Sciences, Beijing 100049, China\\
		$^b$School of Physical Sciences, University of Chinese Academy of Sciences, Beijing 100049, China}
\end{center}

\vspace{1.5cm}

\begin{abstract}
In a class of neutrino mass models with modular flavor symmetries, it has been observed that CP symmetry is preserved at the fixed point (or stabilizer) of the modulus parameter $\tau = {\rm i}$, whereas significant CP violation emerges within the neighbourhood of this stabilizer. In this paper, we first construct a viable model with the modular $A^\prime_5$ symmetry, and explore the phenomenological implications for lepton masses and flavor mixing. Then, we introduce explicit perturbations to the stabilizer at $\tau = {\rm i}$, and present both numerical and analytical results to understand why a small deviation from the stabilizer leads to large CP violation. As low-energy observables are very sensitive to the perturbations to model parameters, we further demonstrate that the renormalization-group running effects play an important role in confronting theoretical predictions at the high-energy scale with experimental measurements at the low-energy scale.

\end{abstract}

\def\thefootnote{\arabic{footnote}}
\setcounter{footnote}{0}

\newpage

\section{Introduction}

The experimental discovery of neutrino oscillations indicates that neutrinos are actually massive and leptonic flavor mixing is significant, but both the origin of neutrino masses and the leptonic flavor structure are largely unknown at present~\cite{Xing:2011zza,Xing:2019vks}. One of the simplest ways to accommodate tiny neutrino masses is to extend the standard model (SM) by introducing three right-handed neutrino singlets $N^{}_{i {\rm R}}$ (for $i = 1,2,3$) and implementing the canonical seesaw mechanism~\cite{Minkowski, Yanagida, Gell-Mann, Glashow, Mohapatra}, which attributes the smallness of three ordinary neutrino masses to the largeness of three heavy Majorana neutrino masses. On the other hand, the leptonic flavor mixing can be accounted for by imposing some discrete flavor symmetries on the seesaw model. In these models, a number of scalar fields transforming nontrivially under the flavor symmetry group have to be introduced, and the flavor structures of lepton mass matrices depend much on the assignments of the relevant fields into the representations of the symmetry group and the choices of the vacuum expectation values (vev's) of the scalar fields~\cite{Altarelli:2010gt, Ishimori:2010au, King:2013eh, King:2014nza}.

Recently, the modular symmetry has been suggested as a possible solution to the flavor mixing puzzle~\cite{Feruglio:2017spp}. In this framework, the Yukawa couplings in the leptonic sector turn out to be modular forms, which are holomorphic functions of the complex modulus $\tau$ and transform as irreducible representations of finite modular groups $\Gamma^{}_{N}$ (with $N \geq 2$ being positive integers). Once the modulus $\tau$ acquires its vev, the Yukawa couplings are determined and thus leptonic flavor mixing pattern is obtained. Finite modular groups, such as $\Gamma^{}_{2} \simeq S^{}_{3}$~\cite{Kobayashi:2018vbk,Kobayashi:2018wkl,Kobayashi:2019rzp}, $\Gamma^{}_{3} \simeq A^{}_{4}$~\cite{Feruglio:2017spp,Kobayashi:2018scp,deAnda:2018ecu,Okada:2018yrn,Okada:2019uoy,Ding:2019zxk,Asaka:2019vev,Zhang:2019ngf,Kobayashi:2019gtp,Wang:2019xbo,Okada:2020rjb,Yao:2020qyy,Chen:2021zty}, $\Gamma^{}_{4} \simeq S^{}_{4}$~\cite{Penedo:2018nmg,Novichkov:2018ovf,Kobayashi:2019mna,Wang:2019ovr,Wang:2020dbp,Zhao:2021jxg},  $\Gamma^{}_{5} \simeq A^{}_{5}$~\cite{Novichkov:2018nkm,Ding:2019xna,Criado:2019tzk}, $\Gamma^{}_{7} \simeq {\rm PSL}(2,Z^{}_7)$~\cite{Ding:2020msi}, and some of their double coverings $\Gamma^\prime_3 \simeq A^\prime_4$~\cite{Liu:2019khw,Lu:2019vgm}, $\Gamma^\prime_4 \simeq S^\prime_4$~\cite{Novichkov:2020eep,Liu:2020akv} and $\Gamma^\prime_5 \simeq A^\prime_5$~\cite{Wang:2020lxk,Yao:2020zml}, have been extensively investigated in the previous literature.

In the bottom-up approach to the modular-invariant flavor models, the modulus $\tau$ is usually treated as a free parameter, which can be pinned down by fitting it together with other model parameters to experimental observations. In contrast, in the top-down approach, it can be fixed by the modulus stabilization via the minimum of the supergravity scalar potential~\cite{Kobayashi:2019xvz} or the flux compactifications~\cite{Ishiguro:2020tmo}. Interestingly, residual symmetries have been found after the spontaneous breaking of the global modular symmetry at some special values of $\tau$, which are called fixed points or stabilizers~\cite{Novichkov:2018ovf,Novichkov:2018yse,deMedeirosVarzielas:2020kji}. If $\tau$ is exactly located at one of the stabilizers, it seems unlikely to realize viable lepton flavor models by using only one modular symmetry~\cite{deMedeirosVarzielas:2019cyj,King:2019vhv} or a common value of $\tau$ in both the charged-lepton and neutrino sectors~\cite{Novichkov:2018yse,Gui-JunDing:2019wap}. One can alternatively construct modular-invariant models with the modulus close to the stabilizers, with which the strong hierarchy of charged-lepton masses can be successfully realized~\cite{Okada:2020ukr,Feruglio:2021dte}.

As pointed out in Ref.~~\cite{Novichkov:2019sqv}, if the generalized CP (gCP) symmetry is further imposed on the modular-invariant model, then both CP and modular symmetries are spontaneously broken by the vev of $\tau$. Moreover, all the stabilizers are CP conserving, while a small deviation of the modulus from them may result in large CP violation~\cite{Novichkov:2019sqv}. Motivated by these observations, we propose in this paper a feasible lepton flavor model with the gCP symmetry and the $\Gamma^\prime_5 \simeq A^\prime_5$ symmetry. A salient feature of our model is the prediction for a nearly-maximal CP violation when $\tau \approx {\rm i}$ holds, whereas CP symmetry is preserved at $\tau = {\rm i}$. It is worth mentioning that a modular-invariant model with large CP violation around the stabilizer $\tau = {\rm i}$ of the modular $A^{}_4$ group has been constructed in Ref.~\cite{Okada:2020brs}. Nevertheless, the question why the large CP violation can be generated with a modulus close to $\tau ={\rm i}$ remains to be answered. To this end, starting with the model at the stabilizer $\tau = {\rm i}$, we introduce explicit perturbations to the modulus and other model parameters, and derive the approximate analytical formulas of neutrino masses, mixing angles and CP-violating phases. Through these analytical calculations, we can see that the nearly-maximal CP violation mainly comes from the large ratio between the real and imaginary parts of $\tau$. In addition, theoretical predictions for some low-energy observables are very sensitive to the perturbations to model parameters when $\tau$ is approaching ${\rm i}$. Such findings also imply that the renormalization-group (RG) running effects may have remarkable impact on the flavor parameters. Therefore, we also take into account the RG running effects and verify that this is indeed the case.

The remaining part of this paper is structured as follows. In Sec.~\ref{sec:modular} we give a brief introduction to the modular symmetry and the modular $\Gamma^\prime_5 \simeq A^\prime_5$ group in order to establish our notations. Then the concrete model for lepton masses and flavor mixing based on the modular $A^\prime_5$ symmetry is constructed in Sec.~\ref{sec:model}. The phenomenological implications for low-energy observables are studied in Sec.~\ref{sec:num} in a completely numerical way, while the analytical analysis of the perturbations to the model with the stabilizer $\tau = {\rm i}$ is performed in Sec.~\ref{sec:ana}. Radiative corrections to the mixing parameters via the RG running are discussed in Sec.~\ref{sec:RGE}. We summarize our main results in Sec.~\ref{sec:sum}. Finally, the basic properties of the finite modular group $\Gamma^{\prime}_5 \simeq A^\prime_5$ are presented in Appendix~\ref{app:A}.

\section{Modular Symmetry}\label{sec:modular}
In this section, we briefly review some basic knowledge about modular symmetries and the modular $\Gamma^\prime_5 \simeq A^\prime_{5}$ group in order to establish our notations and set up the framework for later discussions. The modular group is isomorphic to the special linear group ${\rm SL}(2,\mathbb{Z})$, which is defined as~\cite{Feruglio:2017spp}
\begin{eqnarray}
\Gamma \equiv \left\{\left(\begin{matrix}
a && b \\
c && d
\end{matrix}\right) \bigg| a, b, c, d \in \mathbb{Z}\; ,~ a d - b c = 1 \right\} \; .
% (2.1)
\label{eq:modgr}
\end{eqnarray}
There are in total three generators $S$, $T$ and $R$ for the modular group $\Gamma$, the matrix representations of which are given by
\begin{eqnarray}
S = \left(\begin{matrix}
0 && 1 \\
-1 && 0
\end{matrix}\right) \; , \quad T = \left(\begin{matrix}
1 && 1 \\
0 && 1
\end{matrix}\right) \; , \quad R = \left(\begin{matrix}
-1 && 0 \\
0 && -1
\end{matrix}\right) \; .
\label{eq:STmatrix}
% (2.2)
\end{eqnarray}
The modular transformations on the modulus $\tau$ and chiral supermultiplets $\chi^{(I)}_{}$ are defined as
\begin{eqnarray}
\gamma: \tau \rightarrow \dfrac{a \tau + b}{c \tau + d} \; , \quad
\chi^{(I)}_{} \rightarrow (c \tau +d )^{-k^{}_I} \rho^{(I)}_{} (\gamma) \chi^{(I)} _{} \; ,
\label{eq:lintran}
% (2.3)
\end{eqnarray}
where $\gamma$ is an element of the modular group $\Gamma$, $k^{}_I$ is the weight of the chiral supermultiplet and $\rho^{(I)}_{}(\gamma)$ denotes the unitary representation matrix of $\Gamma$. For the modular-invariant supersymmetric theories, the action ${\cal S}$ should be unchanged under the modular transformations given in Eq.~(\ref{eq:lintran}). As a consequence, the superpotential ${\cal W}$ has to be modular-invariant and the K\"ahler potential ${\cal K}$ should also remain unchanged up to the K\"ahler transformation. For a phenomenological purpose, in the present paper we consider only the minimal form of the K\"ahler potential, namely,
\begin{eqnarray}
{\cal K}(\tau, \bar{\tau}, \chi, \bar{\chi})=-h \log (-{\rm i} \tau+{\rm i} \bar{\tau})+\sum_{I} \frac{\left|\chi^{(I)}_{}\right|^{2}_{}}{(-{\rm i} \tau+{\rm i} \bar{\tau})^{k_{I}}} \; , \nonumber
\label{eq:kahler}
\end{eqnarray}
where $h$ is a positive constant. From Eq.~(\ref{eq:lintran}) one can easily observe that the transformations on $\tau$ induced by $\gamma$ and $-\gamma$ are actually the same, while the matter fields $\chi^{(I)}_{}$ are generally allowed to transform nontrivially under $R$. Therefore, one should consider $\Gamma$ rather than $\overline{\Gamma} \equiv\ \Gamma/\{\mathbb{I}, -\mathbb{I}\}$ as the symmetry group in such theories. In this case, we can introduce the double covering of finite modular groups $\Gamma^\prime_N \equiv \Gamma/\Gamma(N)$ with $\Gamma(N)$ being the principal congruence subgroups of the modular group $\Gamma$, e.g., $\Gamma^{\prime}_{3} \simeq A^{\prime}_4$, $\Gamma^{\prime}_{4} \simeq S^{\prime}_4$ and $\Gamma^{\prime}_{5} \simeq A^{\prime}_5$.

The modular form $f(\tau)$ of level $N$ and weight $k$ is a holomorphic function of $\tau$, which transforms under $\Gamma(N)$ as
\begin{eqnarray}
f\left(\gamma\tau\right)=(c \tau+d)^{k} f(\tau) \; , \quad \gamma \in \Gamma(N) \; ,
\label{eq:modform}
%     (2.4)
\end{eqnarray}
where $k \geq 0$ is an integer. As has been proved in Ref.~\cite{Liu:2019khw}, for a given modular space ${\cal M}^{}_k\left[\Gamma(N)\right]$, the modular forms can always be decomposed into several multiplets that transform as irreducible unitary representations of $\Gamma^{\prime}_{N}$. To be more precise, we can always find a proper basis of the modular space ${\cal M}^{}_k\left[\Gamma(N)\right]$ such that a modular multiplet $Y^{(k)}_{\bf r} = (f^{}_1(\tau),f^{}_2(\tau), \cdots)^{\rm T}_{}$ in the representation ${\bf r}$ satisfies the following equation
\begin{eqnarray}
Y^{(k)}_{\bf r}(\gamma\tau) = (c\tau+d)^k_{}\rho^{}_{\bf r}(\gamma)Y^{(k)}_{\bf r}(\tau)\;, \quad \gamma \in \Gamma\; ,
\label{eq:modformtran}
%     (2.5)
\end{eqnarray}
where $\rho^{}_{\bf r}(\gamma)$ is the identity matrix for $\gamma \in \Gamma(N)$, therefore it is essentially the representation matrix of the quotient group $\Gamma^\prime_N \equiv \Gamma/\Gamma(N)$. For $\Gamma^\prime_5 \simeq A^\prime_5$ in question, the group structure and modular forms with weights from one to six have been analyzed in detail in Ref.~\cite{Wang:2020lxk}. Hence we just recapitulate the key points relevant for the following discussions. The $A^{\prime}_{5}$ group has 120 elements, which can be produced by three generators $S$, $T$ and $R$ satisfying the identities
\begin{eqnarray}
S^2_{}=R \; , \quad (ST)^3_{} = \mathbb{I} \; , \quad T^5_{} = \mathbb{I} \; , \quad R^2_{} = \mathbb{I} \; , \quad RT =TR \; .
\label{eq:genedou}
%     (2.6)
\end{eqnarray}
There are nine distinct irreducible representations for $A^\prime_5$, where the representations ${\bf 1}$, ${\bf 3}$, ${\bf 3}^{\prime}_{}$, ${\bf 4}$ and ${\bf 5}$ with $R$ represented by the unit matrix coincide with those for $A^{}_{5}$, whereas $\widehat{\bf 2}$, $\widehat{\bf 2}^{\prime}_{}$, $\widehat{\bf 4}$ and $\widehat{\bf 6}$ are unique for $A^\prime_5$ with $R$ represented by the minus unit matrix. The irreducible representation matrices of $S$, $T$ and $R$, as well as the decomposition rules of the Kronecker products relevant for the present work, can be found in Appendix~\ref{app:A}.

The modular forms $Y^{(1)}_{\widehat{\bf 6}}(\tau)$ with the lowest nontrivial weight $k=1$ can be expressed as the linear combinations of six basis vectors $\widehat{e}^{}_{i}$ (for $i = 1 , \cdots , 6$) in the modular space ${\cal M}^{}_1[\Gamma(5)]$, whose explicit forms as well as Fourier expansions are given in Appendix~\ref{app:A}. Six components of $Y^{(1)}_{\widehat{\bf 6}}(\tau)$ will be denoted as $Y^{}_i(\tau)$ (for $i = 1, 2, \cdots , 6$), i.e.,
\begin{eqnarray}
Y^{(1)}_{\widehat{\bf 6}} =
\left(\begin{matrix}
Y^{}_1 \\
Y^{}_2 \\
Y^{}_3 \\
Y^{}_4 \\
Y^{}_5 \\
Y^{}_6 \\
\end{matrix}\right)
=\left(\begin{matrix}
\widehat{e}^{}_{1}-3 \, \widehat{e}^{}_6 \\
5\sqrt{2} \, \widehat{e}^{}_2 \\
10 \, \widehat{e}^{}_3 \\
10 \, \widehat{e}^{}_4 \\
5\sqrt{2} \, \widehat{e}^{}_5 \\
-3 \, \widehat{e}^{}_1-\widehat{e}^{}_6 \\
\end{matrix}\right) \; ,
\label{eq:Y_16}
%     (2.7)
\end{eqnarray}
where the argument $\tau$ of all relevant functions has been suppressed. In addition, we write down the modular forms that will be used for the model building in Sec.~\ref{sec:model}. For weight two, we have
\begin{eqnarray}
Y^{(2)}_{{\bf 3}^{\prime}_{}} = \dfrac{1}{2} \left(
\begin{array}{c}
\sqrt{6}\left(Y^2_1-2Y^{}_1Y^{}_6-Y^{2}_6\right) \\
-\sqrt{3} Y^{}_3 (Y^{}_1+Y^{}_6) \\
\sqrt{3} Y^{}_4 (Y^{}_1-Y^{}_6) \\
\end{array}\right) \; , \quad Y^{(2)}_{\bf 5} =\dfrac{1}{2}\left(
\begin{array}{c}
\sqrt{2}\left(Y^2_1+Y^2_6 \right) \\
2\sqrt{6}Y^{}_2 \left( 2Y^{}_1+Y^{}_6\right) \\
\sqrt{3}Y^{}_3\left( Y^{}_6-3Y^{}_1\right) \\
\sqrt{3}Y^{}_4 \left( Y^{}_1+3Y^{}_6\right) \\
2\sqrt{6}Y^{}_5 \left(2Y^{}_6-Y^{}_1\right) \\
\end{array}\right) \; .
\label{eq:Y2}
%     (2.8)
\end{eqnarray}
For weight three, we will use
\begin{eqnarray}
Y^{(3)}_{\widehat{\bf 6},1} &=&  -\dfrac{\sqrt{3}}{2}
\left(
\begin{array}{c}
5 Y_1^3-12 Y_1^{2}Y_6^{}-11 Y_1^{}Y_6^{2}-2 Y_6^3 \\
-2 Y_2^{} \left(Y_1^2-5 Y_1^{}Y_6^{}-2 Y_6^2\right) \\
Y_3^{} \left(Y_1^{}+Y^{}_6\right)\left(Y^{}_1+2Y^{}_6\right) \\
Y_4^{} \left(Y_1^{}-Y^{}_6\right)\left(2Y^{}_1-Y^{}_6\right) \\
2 Y_5^{} \left(2 Y_1^2-5 Y_1^{}Y_6^{}-Y_6^2\right) \\
2 Y_1^3 - 11 Y_1^2 Y_6^{} + 12 Y_1^{} Y_6^2 + 5 Y_6^3 \\
\end{array}
\right) \; , \nonumber
\end{eqnarray}
\begin{eqnarray}
Y^{(3)}_{\widehat{\bf 6},2} &=& -\dfrac{\sqrt{3}}{2}
\left(
\begin{array}{c}
3 Y_1^3-9 Y_1^{2}Y_6^{}-Y_1^{}Y_6^{2}+Y_6^3 \\
-2 Y_2^{} \left(2 Y_1^2-2 Y_1^{}Y_6^{}-Y_6^2\right) \\
2 Y_1^2 Y_3^{} \\
2 Y_4^{} Y_6^2 \\
2 Y_5^{} \left(Y_1^2-2 Y_1^{}Y_6^{}-2 Y_6^2\right) \\
-Y_1^3-Y_1^{2}Y_6^{}+9 Y_1^{}Y_6^{2}+3 Y_6^3 \\
\end{array}
\right) \; .
%     (2.9)
\end{eqnarray}
For weight four, two modular forms $Y^{(4)}_{\bf 1}$ and $Y^{(4)}_{\bf 3}$ are involved, namely,
\begin{eqnarray}
Y^{(4)}_{\bf 1} &=&  -\sqrt{2}\left(Y^4_1-3Y^{3}_{1}Y^{}_{6}-Y^2_{1}Y^2_{6}+3Y^{}_1Y^{3}_{6}+Y^{4}_{6}\right) \; , \nonumber \\
Y^{(4)}_{\bf 3} &=&   \frac{\sqrt{3}}{4}
\left(
\begin{array}{c}
\left(Y_1^2+Y_6^2\right)\left(7 Y_1^2-18  Y_1^{} Y_6^{}-7 Y_6^2\right) \\
Y_2^{} \left(13 Y_1^3-3 Y_1^{2} Y_6-29 Y_1^{} Y_6^2 -9
Y_6^3\right) \\
-Y_5^{} \left(9 Y_1^3-29 Y_1^2 Y_6^{} +3 Y_1^{} Y_6^2 +13
Y_6^3\right) \\
\end{array}
\right) \; .
\label{eq:Y4_1}
% (2.10)
\end{eqnarray}

As we have mentioned, once the modulus $\tau$ acquires its vev, the modular symmetry will be spontaneously broken down. However, there are some special values of $\tau$, which are stabilizers~\cite{Novichkov:2018ovf} and keep unchanged under the transformations induced by one or more generators of the modular group. Consequently, the global modular symmetry is only partially broken, and some residual symmetries are left in the theory. Consider  the fundamental domain ${\cal G}^{}_{}$ of the modular group $\Gamma$, which is defined as
%%%%%%%%%%%%%%%%%%%%%%%%%%%%%%%%%%%   Table 1 %%%%%%%%%%%%%%%%%%%%%%%%%%%%%
\begin{table}[t!]
	\centering
	\vspace{0.1cm}
	\caption{The charge assignment of the chiral superfields under the ${\rm SU(2)^{}_{\rm L}}$ gauge symmetry and the modular $A^{\prime}_{5}$ symmetry in our model, with the corresponding weights listed in the last row.}\vspace{0.5cm}
	\begin{tabular}{ccccccc}
		\toprule
		& $\widehat{L}^{}_e$ & $\widehat{L}^{}_{\mu\tau}$ & $\widehat{E}^{\rm C}_1$  & $\widehat{E}^{\rm C}_{23}$ &  $\widehat{N}^{\rm C}_{}$  & $\widehat{H}^{}_{\rm u},\widehat{H}^{}_{\rm d}$ \\
		\midrule
		{\rm SU(2)} & 2 &2 & 1 & 1  & 1 & 2 \\
		$A^{\prime}_{5}$ & \bf{1} & $\widehat{\bf 2}$ & \bf{1}  & $\widehat{\bf 2}$ & ${\bf 3}^{\prime}_{}$ & \bf{1} \\
		$-k^{}_I$ & 1 & 2 & $-1$ & 2 & 1 & 0 \\
		\bottomrule
		\vspace{0.3cm}
	\end{tabular}
	\label{table:caseb}
\end{table} %%%%%%%%%%%%%%%%%%%%%%%%%%%%%%%%%%%%%%%%%%%%%%%%%%%%%%%%%%%%%%%%%%%%%%%%%%%%%%%%%%
\begin{eqnarray}
{\cal G} = \left\{ \tau \in \mathbb{C}: \quad {\rm Im}\,\tau > 0, \;  -0.5 \leq {\rm Re}\,\tau \leq 0.5, \; |\tau| \geq 1 \right\} \; .
\label{eq:fundo}
% (2.11)
\end{eqnarray}

There are three kinds of stabilizers which are not related to each other by modular transformations in this fundamental domain, which are
\begin{itemize}
	\item $\tau^{}_{\rm C} = {\rm i}$, which is invariant under $S$;
	\item $\tau^{}_{\rm L} = -1/2+{\rm i}\sqrt{3}/2$, invariant under $ST$;
	\item $\tau^{}_{\rm T} = {\rm i}\infty$, invariant under $T$.
\end{itemize}
Actually, there is an additional stabilizer $\tau^{}_{\rm R} = 1/2+{\rm i}\sqrt{3}/2$, which, however, can be obtained from $\tau^{}_{\rm L}$ by the $T$ transformation. In the present paper, we concentrate on the stabilizer $\tau^{}_{\rm C} = {\rm i}$, which is invariant under the transformation corresponding to $S$, i.e., $\tau \rightarrow -1/\tau$. Furthermore, it is straightforward to verify that the residual symmetry at $\tau = {\rm i}$ is $Z^{S}_4 = \{\mathbb{I}, S, R, RS\}$, where $\mathbb{I}$ stands for the identity element.

\section{Modular $A^\prime_5$ Model}\label{sec:model}
Now we are ready to construct a concrete model based on the modular $A^\prime_5$ symmetry. To begin with, we need to assign properly the weights and irreducible representations to all the superfields under the modular group $A^{\prime}_{5}$.  Different from most of the previous models, the superfields of left-handed lepton doublets $\widehat{L}$ will not be arranged as a triplet of $A^{\prime}_5$ in our model. Instead, we take $\widehat{L}^{}_e \sim {\bf 1}$ and $\widehat{L}^{}_{\mu\tau} \equiv (\widehat{L}^{}_{\mu}, \widehat{L}^{}_{\tau})^{\rm T}_{} \sim \widehat{\bf 2}$. Correspondingly, the superfields of right-handed charged leptons are also put into a singlet and a doublet of $A^\prime_5$, i.e., $\widehat{E}^{\rm C}_{1} \sim {\bf 1}$ and  $\widehat{E}^{\rm C}_{23} \equiv (\widehat{E}^{\rm C}_{2}, \widehat{E}^{\rm C}_{3})^{\rm T} \sim \widehat{\bf 2}$. Since the two-dimensional modular forms exist only with the odd weights, the flavor structure of the charged-lepton mass matrix $M^{}_l$ could be highly constrained. As we shall show soon, $M^{}_l$ in our model is restricted to be block-diagonal. In the neutrino sector, the superfields of three right-handed neutrinos $N^{\rm C}_{}$ are set to be ${\bf 3}^\prime_{}$ of $A^\prime_{5}$. In addition, two Higgs doublets $\widehat{H}^{}_{\rm u}$ and $\widehat{H}^{}_{\rm d}$ are both assumed to be in the trivial one-dimensional irreducible representation. All these charge assignments of the superfields in our model are listed in Table~{\ref{table:caseb}.

With the representations and weights of the supermultiplets in Table~{\ref{table:caseb}, one can write down the superpotentials relevant for lepton masses
\begin{eqnarray}
		{\cal W}^{}_l &=& \xi^{}_{1} \left[\widehat{L}^{}_e \widehat{E}^{\rm C}_{1} \right]^{}_{\bf 1} \widehat{H}^{}_{\rm d}  + \xi^{}_{2} \left[\left(\widehat{L}^{}_{\mu\tau} \widehat{E}^{\rm C}_{23}\right)^{}_{\bf 3}Y^{(4)}_{\bf 3} \right]^{}_{\bf 1} \widehat{H}^{}_{\rm d} + \xi^{}_{3} \left[\left(\widehat{L}^{}_{\mu\tau} \widehat{E}^{\rm C}_{23}\right)^{}_{\bf 1}Y^{(4)}_{\bf 1} \right]^{}_{\bf 1} \widehat{H}^{}_{\rm d} \nonumber \; ,
		\\
		{\cal W}^{}_{\rm D} &=& g^{}_{1}\left[\left(\widehat{L}^{}_e \widehat{N}^{\rm C}_{}\right)^{}_{{\bf 3}^{\prime}_{}} Y^{(2)}_{{\bf 3}^{\prime}_{}} \right]^{}_{\bf 1} \widehat{H}^{}_{\rm u} + g^{}_{2}\left[\left( \widehat{L}^{}_{\mu\tau} \widehat{N}^{\rm C}_{}\right)^{}_{\widehat{\bf 6}} Y^{(3)}_{\widehat{\bf 6},1} \right]^{}_{\bf 1} \widehat{H}^{}_{\rm u}+g^{}_{3}\left[\left( \widehat{L}^{}_{\mu\tau} \widehat{N}^{\rm C}_{}\right)^{}_{\widehat{\bf 6}} Y^{(3)}_{\widehat{\bf 6},2} \right]^{}_{\bf 1} \widehat{H}^{}_{\rm u} \nonumber \; ,
		\\
		{\cal W}^{}_{\rm R} &=& \frac{1}{2}\Lambda \left[\left( \widehat{N}^{\rm C}_{} \widehat{N}^{\rm C}_{}\right)^{}_{\bf 5} Y^{(2)}_{\bf 5}\right]^{}_{\bf 1} \; .
		\label{eq:superpB}
		% (3.1)
		\end{eqnarray}
Implementing the Kronecker product rules for $A^\prime_{5}$ collected in Appendix~\ref{app:A}, we arrive at the explicit form of the charged-lepton mass matrix
\begin{eqnarray}
M^{}_l &=& \dfrac{v^{}_{\rm d}}{\sqrt{2}}\left(
		\begin{matrix}
		\xi^{}_1 && 0 && 0\\
		0 && \dfrac{\sqrt{3}}{3}\xi^{}_{2}\left(Y^{(4)}_{\bf 3}\right)^{}_{2}
		&& \xi^{}_{2}\left[\dfrac{\sqrt{6}}{6}\left(Y^{(4)}_{\bf 3}\right)^{}_{1}-\dfrac{\sqrt{2}}{2}\widetilde{\xi}Y^{(4)}_{\bf 1}\right] \\
		0 && \xi^{}_2 \left[\dfrac{\sqrt{6}}{6}\left(Y^{(4)}_{\bf 3}\right)^{}_{1}+\dfrac{\sqrt{2}}{2}\widetilde{\xi}Y^{(4)}_{\bf 1}\right] && -\dfrac{\sqrt{3}}{3}\xi^{}_{2}\left(Y^{(4)}_{\bf 3}\right)^{}_{3}
		\end{matrix}\right)^\ast_{} \; , \label{eq:Ml}
%     (3.2)
\end{eqnarray}
with $\xi^{}_3/\xi^{}_2 \equiv \widetilde{\xi}$, the Dirac neutrino mass matrix
\begin{eqnarray}
M^{}_{\rm D} &=& \dfrac{\sqrt{6}g^{}_1 v^{}_{\rm u}}{12} \left[\rule{0cm}{1.3cm}\right. \left(\begin{matrix}
		2\left(Y^{(2)}_{\bf 3^{\prime}_{}}\right)^{}_1 && 2\left(Y^{(2)}_{\bf 3^{\prime}_{}}\right)^{}_3 && 2\left(Y^{(2)}_{\bf 3^{\prime}_{}}\right)^{}_2 \\
		0 && 0 && 0 \\
		0 && 0 && 0 \\
		\end{matrix}\right)  \nonumber \\
		&& +\widetilde{g}^{}_2\left(\begin{matrix}
		0 &&  0 && 0 \\
		-\sqrt{2} \left(Y^{(3)}_{\widehat{\bf 6},1}\right)^{}_4 && -\sqrt{2} \left(Y^{(3)}_{\widehat{\bf 6},1}\right)^{}_2 && \left(Y^{(3)}_{\widehat{\bf 6},1}\right)^{}_1-\left(Y^{(3)}_{\widehat{\bf 6},1}\right)^{}_6 \\
		-\sqrt{2}\left(Y^{(3)}_{\widehat{\bf 6},1}\right)^{}_3 && \left(Y^{(3)}_{\widehat{\bf 6},1}\right)^{}_1+\left(Y^{(3)}_{\widehat{\bf 6},1}\right)^{}_6 && -\sqrt{2}\left(Y^{(3)}_{\widehat{\bf 6},1}\right)^{}_5 \\
\end{matrix}\right) \nonumber \\
&& +\widetilde{g}^{}_3 \left(\begin{matrix}
		0 &&  0 && 0 \\
		-\sqrt{2} \left(Y^{(3)}_{\widehat{\bf 6},2}\right)^{}_4 && -\sqrt{2} \left(Y^{(3)}_{\widehat{\bf 6},2}\right)^{}_2 && \left(Y^{(3)}_{\widehat{\bf 6},2}\right)^{}_1-\left(Y^{(3)}_{\widehat{\bf 6},2}\right)^{}_6 \\
		-\sqrt{2}\left(Y^{(3)}_{\widehat{\bf 6},2}\right)^{}_3 && \left(Y^{(3)}_{\widehat{\bf 6},2}\right)^{}_1+\left(Y^{(3)}_{\widehat{\bf 6},2}\right)^{}_6 && -\sqrt{2}\left(Y^{(3)}_{\widehat{\bf 6},2}\right)^{}_5 \\
\end{matrix}\right) \left. \rule{0cm}{1.3cm}\right]^\ast_{}  \; , \label{eq:MD}
%     (3.3)
\end{eqnarray}
with $g^{}_2/g^{}_1 \equiv \widetilde{g}^{}_2$ and $g^{}_3/g^{}_1 \equiv \widetilde{g}^{}_3$, and the Majorana mass matrix of right-handed neutrinos
\begin{eqnarray}
M^{}_{\rm R} &=& \dfrac{\Lambda}{4} \left(\begin{matrix}
		2\left(Y^{(2)}_{\bf 5}\right)^{}_1 && -\sqrt{3}\left(Y^{(2)}_{\bf 5}\right)^{}_4 && -\sqrt{3}\left(Y^{(2)}_{\bf 5}\right)^{}_3 \\
		-\sqrt{3}\left(Y^{(2)}_{\bf 5}\right)^{}_4 && \sqrt{6}\left(Y^{(2)}_{\bf 5}\right)^{}_2 && -\left(Y^{(2)}_{\bf 5}\right)^{}_1 \\
		-\sqrt{3}\left(Y^{(2)}_{\bf 5}\right)^{}_3 && -\left(Y^{(2)}_{\bf 5}\right)^{}_1 && \sqrt{6}\left(Y^{(2)}_{\bf 5}\right)^{}_5 \\
		\end{matrix} \right)^\ast_{} \; , \label{eq:MR}
		% (3.4)
\end{eqnarray}
where $(Y^{(k)}_{\bf r})^{}_i$ denotes the $i$-th element in the multiplet $Y^{(k)}_{\bf r}$. Given $M^{}_{\rm D}$ and $M^{}_{\rm R}$, we can immediately derive the effective neutrino mass matrix from the seesaw formula, i.e., $M^{}_\nu \approx - M^{}_{\rm D}M^{-1}_{\rm R}M^{\rm T}_{\rm D}$. Hence the lepton mass spectra and flavor mixing parameters can be extracted from the charged-lepton mass matrix $M^{}_l$ and the effective neutrino mass matrix $M^{}_\nu$.

Apart from the modulus $\tau$, the lepton mass matrices involve three extra parameters $\widetilde{\xi}^{}$, $\widetilde{g}^{}_2$ and $\widetilde{g}^{}_3$, which are in general complex. All these complex parameters may contribute to leptonic CP violation. Following Ref.~\cite{Novichkov:2019sqv}, we further impose gCP symmetry on our modular $A^\prime_5$ model such that the modulus $\tau$ becomes the only source of CP violation. More explicitly, the gCP transformation of the superfield $\chi^{(I)}_{}(x)$ is
\begin{eqnarray}
		\chi^{(I)}_{}(x) \stackrel{\rm CP}{\longrightarrow} X^{}_{\bf r} \overline{\chi}^{(I)}_{}(x^{}_{\rm P}) \; ,
\label{eq:gcpdef}
	% (3.5)
\end{eqnarray}
where $\overline{\chi}^{(I)}_{}(x^{}_{\rm P})$ denotes the conjugate superfield with $x^{}_{\rm P} = (t,-\vec{x})$ and $X^{}_{\bf r}$ represents a unitary matrix acting on the flavor space. For the gCP symmetry to be consistent with the modular symmetry, the following condition should be satisfied
\begin{eqnarray}
X^{}_{\bf{r}} \rho_{\bf{r}}^{*}(\gamma) X_{\bf{r}}^{-1}=\rho_{\bf{r}}^{}(u(\gamma)) \; ,
\label{eq:consist}
	% (3.6)
\end{eqnarray}
where $u(\gamma)$ is an outer automorphism of the modular group.\footnote{It has been mentioned in Ref.~\cite{Novichkov:2020eep} that there are two distinct kinds of outer automorphisms $u(\gamma)$ for the double covering group, corresponding to two different gCP symmetries denoted as ${\rm CP^{}_1}$ and ${\rm CP^{}_2}$, respectively. However, in the basis where the representation matrices of $S$ and $T$ are both symmetric, $X^{}_{\bf r}$ will always be the trivial identity matrix, i.e., $X^{}_{\bf r} = \mathbb{I}$, regardless of whether ${\rm CP}^{}_{1}$ or ${\rm CP}^{}_{2}$ is combined with the $\Gamma^\prime_5$ group.} The consistency condition given in Eq.~(\ref{eq:consist}) indicates that the modulus $\tau$ transforms under CP as
\begin{eqnarray}
\tau \stackrel{\rm CP}{\longrightarrow} -\tau^{\ast}_{} \; .
\label{eq:tauCP}
     % (3.7)
\end{eqnarray}
	 On the other hand, Eq.~(\ref{eq:consist}) should be satisfied for all the elements in the finite modular group, implying $X^{}_{\bf r} = \mathbb{I}$ with $\mathbb{I}$ being the identity element if the representation matrices of both $S$ and $T$ are symmetric~\cite{Novichkov:2019sqv}. Furthermore, as can be seen in Appendix~\ref{app:A}, all the Clebsch-Gordan coefficients are real, so the modular forms $Y^{(k)}_{\bf r}$ will transform under CP as
\begin{eqnarray}
Y_{\bf{r}}^{(k)}(\tau) \stackrel{\mathrm{C P}}{\longrightarrow} Y_{\bf{r}}^{(k)}\left(-\tau^{*}\right)=\left[Y_{\bf{r}}^{(k)}(\tau)\right]^{*} \; .
\label{eq:modformCP}
	% (3.8)
\end{eqnarray}
Therefore, to render the superpotentials invariant under the gCP transformation, we require all the coupling constants in our model to be real. As a result, the whole symmetry of modular and gCP transformations is spontaneously broken down after the modulus $\tau$ gets its vev. However, there are some special values of $\tau$, for which CP symmetry is conserved while the modular symmetry is broken. As pointed out in Ref.~\cite{Novichkov:2019sqv}, these values are located along the imaginary axis (i.e., ${\rm Re}\,\tau = 0$) and the boundary of the fundamental domain ${\cal G}$.

\section{Low-energy Phenomenology}\label{sec:num}
	%%%%%%%%%%%%%%%%%%%% Table 1 %%%%%%%%%%%%%%%%%%%%%%%%%%%%%%%%%%%%%%%%%%%%%%%	
\begin{table}[t]
	\begin{center}
		\vspace{-0.25cm} \caption{The best-fit values, the
			1$\sigma$ and 3$\sigma$ intervals, together with the values of $\sigma^{}_{i}$ being the symmetrized $1\sigma$ uncertainties, for three neutrino mixing angles $\{\theta^{}_{12}, \theta^{}_{13}, \theta^{}_{23}\}$, two neutrino mass-squared differences $\{\Delta m^2_{21}, \Delta m^2_{31}~{\rm or}~\Delta m^2_{32}\}$ and the Dirac CP-violating phase $\delta$ from a global-fit analysis of current experimental data~\cite{Esteban:2020cvm}.} \vspace{0.5cm}
		\begin{tabular}{c|c|c|c|c}
			\hline
			\hline
			Parameter & Best fit & 1$\sigma$ range &  3$\sigma$ range & $\sigma^{}_{i}$ \\
			\hline
			\multicolumn{5}{c}{Normal neutrino mass ordering
				$(m^{}_1 < m^{}_2 < m^{}_3)$} \\ \hline
			%------------------------------------------------------------
			$\sin^2_{}\theta^{}_{12}$
			& $0.304$ & 0.292 --- 0.317 &  0.269 --- 0.343 & 0.0125 \\
			%------------------------------------------------------------
			$\sin^2_{}\theta^{}_{13}$
			& $0.02221$ & 0.02159 --- 0.02289 &  0.02034 --- 0.02430  & 0.00065 \\
			%------------------------------------------------------------
			$\sin^2_{}\theta^{}_{23}$
			& $0.570$  & 0.546 --- 0.588 &  0.407 --- 0.618  & 0.021 \\
			%------------------------------------------------------------
			$\delta/{}^\circ_{}$ &  $195$ & 170 --- 246 &  107 --- 403 & 38 \\
			%------------------------------------------------------------
			$\Delta m^2_{21}/ (10^{-5}~{\rm eV}^2)$ &  $7.42$ & 7.22 --- 7.63 & 6.82 --- 8.04 & 0.205 \\
			%------------------------------------------------------------
			$\Delta m^2_{31}/(10^{-3}~{\rm eV}^2)$ &  $+2.514$ & +2.487 --- +2.542 & +2.431 --- +2.598 & 0.0275 \\\hline
			%%%%%%%%%%%%%%%%%%%%%%%%%%%%%%%%%%%%%%%%%%%%%%%%%%%%%%%%%%%%%
			\multicolumn{5}{c}{Inverted neutrino mass ordering
				$(m^{}_3 < m^{}_1 < m^{}_2)$} \\ \hline
			%------------------------------------------------------------
			$\sin^2_{}\theta^{}_{12}$
			& $0.304$ & 0.292 --- 0.317 &  0.269 --- 0.343 & 0.0125\\
			%------------------------------------------------------------
			$\sin^2_{}\theta^{}_{13}$
			& $0.02240$ & 0.02178 --- 0.02302 &  0.02053 --- 0.02436 & 0.00062 \\
			%------------------------------------------------------------
			$\sin^2_{}\theta^{}_{23}$
			& $0.575$  & 0.554 --- 0.592 &  0.411 --- 0.621 & 0.019  \\
			%------------------------------------------------------------
			$\delta/{}^\circ_{}$ &  $286$ & 254 --- 313 &  192 --- 360 & 29.5 \\
			%------------------------------------------------------------
			$\Delta m^2_{21}/ (10^{-5}~{\rm eV}^2)$ &  $7.42$ & 7.22 --- 7.63 & 6.82 --- 8.04 & 0.205 \\
			%------------------------------------------------------------
			$\Delta m^2_{32}/(10^{-3}~{\rm eV}^2)$ &  $-2.497$ & $-2.525$ --- $-2.469$  & $-2.583$ --- $-2.412$ & $0.028$ \\ \hline\hline
			%%%%%%%%%%%%%%%%%%%%%%%%%%%%%%%%%%%%%%%%%%%%%%%%%%%%%%%%%%%%%
		\end{tabular}
		\label{table:gfit}
	\end{center}
\end{table}
%%%%%%%%%%%%%%%%%%%%%%%%%%%%%%%%%%%%%%%%%%%%%%%%%%%%%%%%%%%%%%%%%
Thanks to the gCP symmetry, only eight real parameters $\{{\rm Re}\,\tau, {\rm Im}\,\tau,\xi^{}_1 v^{}_{\rm d}, \xi^{}_2 v^{}_{\rm d}, \widetilde{\xi}^{}, g^{2}_1 v^2_{\rm u} /\Lambda,\widetilde{g}^{}_2,\widetilde{g}^{}_3\}$ are involved in our model. The allowed regions of model parameters can be found by following the numerical analysis adopted in Ref.~\cite{Wang:2020lxk}, and the basic strategy is summarized as below.
\begin{itemize}
\item First, we explain the experimental results that are used to constrain the parameter space. For the charged-lepton masses, we take the best-fit values $m^{}_{e} = 0.510~{\rm MeV}$, $m^{}_{\mu} = 107.8~{\rm MeV}$ and $m^{}_{\tau} = 1.840~{\rm GeV}$, which are evaluated at the scale $\Lambda^{}_{\rm GUT} \approx 2 \times 10^{16}_{}~{\rm GeV}$ of grand unified theories (GUT) with $\tan\beta \equiv v^{}_{\rm u}/v^{}_{\rm d}=10$ and the supersymmetry breaking scale $m^{}_{\rm SUSY} = 10~{\rm TeV}$ in Ref.~\cite{Antusch:2013jca}. With the help of these charged-lepton masses, one can determine the model parameters $\xi^{}_1 v^{}_{\rm d}$, $\xi^{}_2 v^{}_{\rm d}$ and $\widetilde{\xi}$ if the complex modulus $\tau$ is given. For the neutrino sector, we take the best-fit values, as well as $1\sigma$ and $3\sigma$ ranges, of  two neutrino mass-squared differences $\Delta m^{2}_{21} \equiv m^2_2-m^2_1$ and  $\Delta m^{2}_{31} \equiv m^2_3-m^2_1$ in the normal mass ordering (NO) with $m^{}_1<m^{}_2<m^{}_3$ or $\Delta m^{2}_{21}$ and $\Delta m^{2}_{32} \equiv m^2_3-m^2_2$ in the inverted mass ordering (IO) with $m^{}_3<m^{}_1<m^{}_2$, three flavor mixing angles $\{\theta^{}_{12}, \theta^{}_{13}, \theta^{}_{23}\}$, and the Dirac CP-violating phase $\delta$, from the global-fit analysis by NuFIT 5.0~\cite{Esteban:2020cvm,nufit5.0} without including the atmospheric neutrino data from Super-Kamiokande. All these values are summarized in Table~\ref{table:gfit}.

\item The model predictions for low-energy observables, including charged-lepton and neutrino masses, flavor mixing angles, and CP-violating phases, can be obtained by diagonalizing lepton mass matrices $M^{}_l$ and $M^{}_\nu$. Then, the compatibility between the model predictions and the experimental observations is measured by the $\chi^2$-function, which is constructed as the sum of several one-dimensional functions $\chi^2_{j}$, namely,
\begin{eqnarray}
\chi^2_{}(p^{}_{i}) = \sum^{}_{j}\chi^2_{j}(p^{}_i)\; ,
\label{eq:chi}
%     (3.8)
\end{eqnarray}
where $p^{}_{i} \in \{{\rm Re}\,\tau, {\rm Im}\,\tau,\widetilde{g}^{}_2,\widetilde{g}^{}_3,g^{2}_1 v^2_{\rm u} /\Lambda\}$ stand for the model parameters, and $j$ is summed over the observables $\{\sin^2\theta^{}_{12}, \sin^2\theta^{}_{13}, \sin^2\theta^{}_{23}, \Delta m^2_{21}, \Delta m^2_{31} (\Delta m^2_{32})\}$ in the NO (IO) case. Here we do not include the information of $\delta$ in the $\chi^2_{}$-function due to the weak constraints on $\delta$ from the global-fit results. For $\sin^2\theta^{}_{12}$, $\sin^2\theta^{}_{13}$, $\Delta m^2_{21}$ and $\Delta m^2_{31}$ ($\Delta m^2_{32}$), we make use of the Gaussian approximations
\begin{eqnarray}
\chi^2_{j}(p^{}_{i}) = \left(\frac{q^{}_{j}(p^{}_{i})-q^{\rm bf}_{j}}{\sigma^{}_{j}}\right)^2_{} \; ,
\label{eq:ch2}
%     (3.9)
\end{eqnarray}
where $q^{}_j(p^{}_i)$ denote the model predictions for these observables and $q^{\rm bf}_j$ are their best-fit values from the global analysis in Ref.~\cite{Esteban:2020cvm}. The associated uncertainties $\sigma^{}_j$ are derived by symmetrizing $1\sigma$ uncertainties from the global-fit analysis, as given in Table~\ref{table:gfit}. For $\sin^2_{}\theta^{}_{23}$, we utilize the one-dimensional projection of the $\chi^2$-function provided by Refs.~\cite{Esteban:2020cvm,nufit5.0}. By minimizing the overall $\chi^2_{}$-function in Eq.~(\ref{eq:chi}), we can determine the best-fit values of the model parameters $\{{\rm Re}\,\tau, {\rm Im}\,\tau,\widetilde{g}^{}_2,\widetilde{g}^{}_3, g^{2}_1 v^2_{\rm u} /\Lambda\}$. 
\end{itemize}
%%%%%%%%%%%%%%%%%%%%%%%%%%%%%%% Fig. 1 %%%%%%%%%%%%%%%%%%%%%%%%%%%%%%%%
\begin{figure}[t!]
	\centering		\includegraphics[width=0.99\textwidth]{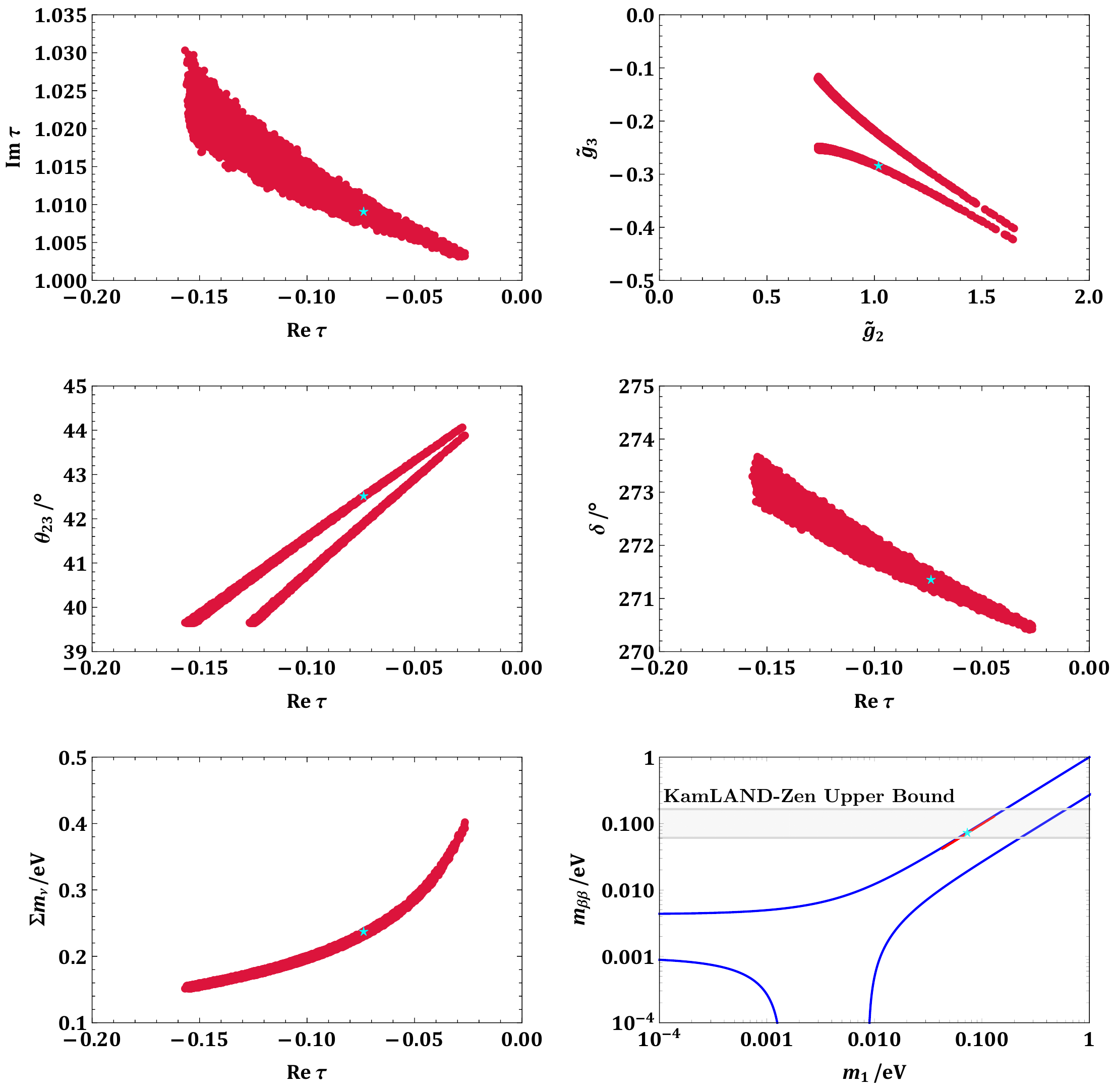} 	\vspace{0.3cm}
	\caption{The allowed parameter space of the model parameters $\{{\rm Re}\,\tau, {\rm Im}\,\tau,\widetilde{g}^{}_2,\widetilde{g}^{}_3\}$ and low-energy observables at the $3\sigma$ level in the NO case, where the cyan stars denote the best-fit values from the $\chi^2$-fit analysis. The gray shaded region in the bottom-right panel represents the upper bound on $m^{}_{\beta\beta}$ from the KamLAND-Zen experiment~\cite{KamLAND-Zen:2016pfg}, and the blue boundary is obtained by using the $3\sigma$ ranges of $\{\theta^{}_{12},\theta^{}_{13}\}$ and $\{\Delta m^{2}_{21},\Delta m^{2}_{31}\}$ from the global-fit analysis.}
	\label{fig:figPara} %% label for entire figure
	\vspace{0.3cm}
\end{figure}
%%%%%%%%%%%%%%%%%%%%%%%%%%%%%%%%%%%%%%%%%%%%%%%%%%%%%%%%%%%%%%%%%%%%%%%

After carrying out the numerical analysis, we find that our model can be compatible with the experimental data at the $3\sigma$ level only in the NO case. In Fig.~\ref{fig:figPara}, the $3\sigma$ allowed parameter space of model parameters as well as the predictions for low-energy observables are shown. Some comments on the numerical results are in order.
\begin{enumerate}
\item In the top-left panel of Fig.~\ref{fig:figPara}, we can observe that the $3\sigma$ allowed value of ${\rm Re}\,\tau$ is larger than $-0.16$, while the upper bound of ${\rm Im}\,\tau$ is around $1.03$. In fact, there should be a duplicate region of $\{{\rm Re}\,\tau,{\rm Im}\,\tau\}$ in the right-half part of the fundamental domain ${\cal G}$ with $0 \leq {\rm Re}\,\tau \leq 0.5$, where the signs of all CP-violating phases are reversed. However, the predicted value of the Dirac CP-violating phase $\delta$ in this region is constrained to be about $90^\circ_{}$, which is lying outside the $3\sigma$ allowed range of $\delta$ from the global-fit analysis. The allowed regions of the other two parameters $\{\widetilde{g}^{}_2,\widetilde{g}^{}_3\}$ in the neutrino sector are shown in the top-right panel of Fig.~\ref{fig:figPara}, where it is worthwhile to mention that the relation $\widetilde{g}^{}_3 = -\widetilde{g}^{}_2/4$ holds approximately, especially for large values of $\widetilde{g}^{}_2$. Some of the low-energy observables are strongly correlated with ${\rm Re}\,\tau$ (or ${\rm Im}\,\tau$). For example, the middle-left panel of Fig.~\ref{fig:figPara} reveals that the value of $\theta^{}_{23}$ increases as ${\rm Re}\,\tau$ decreases. When the value of ${\rm Re}\,\tau$ is approaching zero, $\theta^{}_{23}$ becomes very close to $45^\circ_{}$. Interestingly, the Dirac CP-violating phase $\delta$ shows a similar behavior, i.e., $\delta$ tends to $270^\circ_{}$ when ${\rm Re}\,\tau \rightarrow 0$, implying a nearly-maximal CP violation in the vicinity of $\tau = {\rm i}$. This result seems quite confusing, since the stabilizer $\tau={\rm i}$ should lead to CP conservation. Why does a small deviation from $\tau = {\rm i}$ (i.e., ${\rm Im}\, \tau - 1 < 0.5\%$) give rise to a large CP-violating phase? In the next section, we shall come back to answer this question by performing an analytical analysis with some reasonable approximations. The other mixing angles $\theta^{}_{12}$ and $\theta^{}_{13}$ and two neutrino mass-squared differences are loosely constrained and thus not presented here.

\item The sum of three light neutrino masses $\sum m^{}_\nu = m^{}_1+m^{}_2+m^{}_3$ also depends crucially on the value of ${\rm Re}\,\tau$. As indicated in the bottom-left panel of Fig.~\ref{fig:figPara}, the minimal value of $\sum m^{}_\nu $ predicted in our model is $0.15~{\rm eV}$, which runs into contradiction with the most stringent bound $\sum m^{}_{\nu} < 0.12~{\rm eV}$ from the Planck observations of cosmic microwave background~\cite{Aghanim:2018eyx}. Notice that the limit $\sum m^{}_{\nu} < 0.12~{\rm eV}$ has also been gained earlier in Ref.~\cite{Giusarma:2016phn,Vagnozzi:2017ovm}. However, this upper bound is dependent on the choices of observational data sets and cosmological models. The latest global-fit analysis of absolute neutrino masses yields $\sum m^{}_{\nu} < (0.12 \cdots 0.69)~{\rm eV}$ at the $2\sigma$ level~\cite{Capozzi:2017ipn}, where the ``aggressive" and ``conservative" combinations of different observational data sets are considered. Therefore, our model is still consistent with the ``conservative" cosmological bound on neutrino masses.
\item Two Majorana CP-violating phases $\rho$ and $\sigma$ can also be determined, so we can calculate the effective mass for neutrinoless double-beta decays
\begin{eqnarray}
m^{}_{\beta\beta} \equiv |m^{}_1 \cos^{2}_{}\theta^{}_{13} \cos^{2}_{}\theta^{}_{12}e^{2{\rm i}\rho}_{} + m^{}_2 \cos^{2}_{}\theta^{}_{13} \sin^{2}_{}\theta^{}_{12}e^{2{\rm i}\sigma}_{} + m^{}_3 \sin^2_{}\theta^{}_{13} e^{-2{\rm i}\delta}_{}| \; ,
\label{eq:mbb}
% (4.3)
\end{eqnarray}
where the standard parametrization of leptonic flavor mixing matrix has been taken~\cite{PDG2020}. The numerical result of $m^{}_{\beta\beta}$ is presented in the bottom-right panel of Fig.~\ref{fig:figPara}, where we can find that the predicted range of $m^{}_{\beta\beta}$ is overlapping with the upper bound $m^{\rm upper}_{\beta\beta} = (61\cdots 165)~{\rm meV}$ from the KamLAND-Zen experiment~\cite{KamLAND-Zen:2016pfg}. Hence the next-generation neutrinoless double-beta decay experiments with a sensitivity of $m^{}_{\beta\beta} \sim 15~{\rm meV}$ ~\cite{Dolinski:2019nrj} will be able to make a final verdict on whether our model is ruled out or not.
\end{enumerate}

Based on the $\chi^2$-fit analysis, we find that the minimum $\chi^{2}_{\rm min} = 0.735$ is reached in the NO case with the following best-fit values of the model parameters
\begin{eqnarray}
{\rm Re}\,\tau = -0.0735\;, \quad  {\rm Im}\,\tau = 1.00897 \;, \quad
\widetilde{g}^{}_2 = 1.0199 \;, \quad
\widetilde{g}^{}_3 = -0.286 \; , \quad  g^{2}_1 v^{2}_{\rm u}/\Lambda = 62.35~{\rm meV} \; . \nonumber
\label{eq:bfcaseA}
%     (4.4)
\end{eqnarray}
Combining the above values with charged-lepton masses, we obtain $\xi^{}_1 v^{}_{\rm d}= 0.721~{\rm MeV}$, $ \xi^{}_2 v^{}_{\rm d} = 0.246~{\rm GeV}$ and $\widetilde{\xi}= 3.981$. These best-fit values of model parameters lead to a nearly-degenerate spectrum of three neutrino masses $m^{}_{1} = 72.58~{\rm meV}$, $m^{}_{2} = 73.08~{\rm meV}$, $m^{}_{3} = 88.21~{\rm meV}$, and three mixing angles $\theta^{}_{12} = 33.58^\circ_{}$, $\theta^{}_{13} = 8.58^\circ_{}$ and $\theta^{}_{23} = 42.51^\circ_{}$. Meanwhile, the predictions for three CP-violating phases are $\delta = 271.3^\circ_{}$, $\rho = 88.74^\circ_{}$ and $\sigma = 90.53^\circ_{}$. In addition, the effective mass for neutrinoless double-beta decays is $m^{}_{\beta\beta} = 71.11~{\rm meV}$. All these predictions are readily to be tested in future neutrino oscillation and neutrinoless double-beta decay experiments.

\section{Explicit Perturbations and Analytical Results}\label{sec:ana}

As we have seen in Sec.~\ref{sec:num}, our model predicts a nearly-maximal CP-violating phase $\delta = 271.4^\circ_{}$, given ${\rm Re}\,\tau = -0.0736$ and ${\rm Im}\,\tau = 1.00897$, which is lying in the vicinity of $\tau = {\rm i}$. Since CP violation is completely absent for $\tau = {\rm i}$, it deserves a further investigation to clarify how such significant CP violation is generated. In this section, we attempt to explore the analytical properties of our model in the parameter space, where $\tau$ is close to the stabilizer $\tau = {\rm i}$, by introducing explicit perturbations to the stabilizer and calculating lepton mass spectra and flavor mixing parameters.

Let us first consider the NO case. If $\tau={\rm i}$ exactly holds, with the help of the explicit expressions of basis vectors $\widehat{e}^{}_i$ (for $i = 1,\cdots,6$) shown in Eq.~(\ref{eq:basvec}), we can find
\begin{eqnarray}
\frac{\widehat{e}^{}_2(\rm i)}{\widehat{e}^{}_1(\rm i)} = \frac{\widehat{e}^{}_3(\rm i)}{\widehat{e}^{}_2(\rm i)} =
\frac{\widehat{e}^{}_4(\rm i)}{\widehat{e}^{}_3(\rm i)} =
\frac{\widehat{e}^{}_5(\rm i)}{\widehat{e}^{}_4(\rm i)} =
\frac{\widehat{e}^{}_6(\rm i)}{\widehat{e}^{}_5(\rm i)} = A^{}_0 \; ,
\label{eq:eratio}
% (5.1)
\end{eqnarray}
where $A^{}_0 = \sqrt{\sqrt{5}\phi}-\phi$ with $\phi \equiv (\sqrt{5}+1)/2$. The elements of $Y^{(1)}_{\widehat{\bf 6}}(\rm i)$ can be explicitly written as
\begin{eqnarray}
\begin{array}{lllll}
Y^{}_1({\rm i}) = \widehat{e}^{}_1({\rm i}) (1-3 \, A^5_0 ) \; , &\quad\quad & Y^{}_2({\rm i}) = 5\sqrt{2}\,\widehat{e}^{}_1({\rm i}) A^{}_0 \; , &\quad\quad & Y^{}_3({\rm i}) = 10 \, \widehat{e}^{}_1({\rm i}) A^{2}_0 \; , \\
Y^{}_4({\rm i}) = 10\,\widehat{e}^{}_1({\rm i}) A^{3}_0 \; , &\quad\quad & Y^{}_5({\rm i}) =  5\sqrt{2}\,\widehat{e}^{}_1({\rm i}) A^{4}_0 \; , &\quad\quad & Y^{}_6({\rm i}) = -\widehat{e}^{}_1({\rm i})(3 + A^{5}_0) \; .
\end{array}
\label{eq:Y_16re}
%     (5.2)
\end{eqnarray}
Then the modular forms at $\tau = {\rm i}$ with higher weights can be constructed by using Eq.~(\ref{eq:Y_16re}). Since the stabilizer $\tau={\rm i}$ keeps unchanged under the transformation of $S$, $H^{}_{l} \equiv M^{}_l M^{\dag}_l$ from the charged-lepton sector is invariant under $S$, i.e.,
\begin{eqnarray}
H^{}_l = \rho^{\dag}_l(S) H^{}_l \rho^{}_{l}(S) \; ,
\label{eq:Mltran}
% (5.3)
\end{eqnarray}
where $\rho^{}_l(S) \equiv {\rm Diag}\{1,\rho^{}_{\bf 2}(S)\}$ with $\rho^{}_{\bf 2}(S)$ being the representation matrix of the two-dimensional irreducible representation {\bf 2} of $S$ in the $A^\prime_{5}$ group. The identity in Eq.~(\ref{eq:Mltran}) tells us that a unitary matrix converting $\rho^{}_l(S)$ into its diagonal form can also diagonalize the matrix $H^{}_l$. It is easy to verify that both $\rho^{}_l(S)$ and $H^{}_l$ can be diagonalized by the real orthogonal matrix
\begin{eqnarray}
U^{}_l = \left(
\begin{matrix}
1 & 0 & 0 \\
0 & \cos\theta^{}_l &-\sin\theta^{}_l \\
0 & \sin\theta^{}_l & \cos\theta^{}_l \\
\end{matrix}\right) \; ,
\label{eq:Ul}
% (5.4)
\end{eqnarray}
where $\tan 2\theta^{}_l = \sqrt{(\phi-1)/\phi}$. Two comments on $U^{}_l$ in Eq.~(\ref{eq:Ul}) are helpful. First, substituting $\phi = (\sqrt{5}+1)/2 \approx 1.618$ into Eq.~(\ref{eq:Ul}), we arrive at $\cos\theta^{}_l \approx 0.962$ and $\sin\theta^{}_l \approx 0.273$. Therefore, $U^{}_l$ is roughly an identity matrix, and its contribution to the lepton flavor mixing is insignificant. Second, we have checked that the form of $U^{}_l$ given in Eq.~(\ref{eq:Ul}) holds as a good approximation to the unitary matrix that diagonalizes $H^{}_l$, even if $\tau$ slightly deviates from the stabilizer $\tau = {\rm i}$. It is then safe to assume $U^{}_l$ in Eq.~(\ref{eq:Ul}) to be valid in the vicinity of $\tau = {\rm i}$.

We turn to the neutrino sector. Without loss of generality, one can work in the flavor basis where the charged-lepton mass matrix is diagonal, and redefine the effective neutrino mass matrix as $\widetilde{M}^{}_\nu = U^\dag_l M^{}_\nu U^{}_l$. For $\tau = {\rm i}$, $\widetilde{M}^{}_\nu$ is simply written as
\begin{eqnarray}
\widetilde{M}^{}_\nu = \frac{\widehat{g}^2_1 v^2_{\rm u}}{ {\rm Det}(M^{}_{\rm R})}
\left(\begin{matrix}
50.67 && 0 && -17.19(\widehat{g}^{}_2+4\widehat{g}^{}_3) \\
0 && 0 && 0 \\
-17.19(\widehat{g}^{}_2+4\widehat{g}^{}_3)  && 0 && 5.834(\widehat{g}^{}_2+4\widehat{g}^{}_3)^2_{} \\
\end{matrix}\right) \; ,
\label{eq:mnu0}
% (5.5)
\end{eqnarray}
with $\widehat{g}^{}_1 = g^{}_1 |\widehat{e}^{}_1(\tau)|^2_{}$ and $\widehat{g}^{}_{2,3} = \widetilde{g}^{}_{2,3} |\widehat{e}^{}_1(\tau)|$. It is obvious that the right-handed neutrino mass matrix $M^{}_{\rm R}$ has one zero eigenvalue for $\tau = {\rm i}$, i.e., ${\rm Det}(M^{}_{\rm R}) = 0$ in Eq.~(\ref{eq:mnu0}). The seesaw formula $M^{}_\nu = - M^{}_{\rm D}M^{-1}_{\rm R}M^{\rm T}_{\rm D}$ is not directly applicable for $\tau = {\rm i}$, but anyway we are interested in the region where $\tau = {\rm i}+\epsilon$ with $|\epsilon| \ll 1$. In this case, the determinant of $M^{}_{\rm R}$ is found to be ${\rm Det}(M^{}_{\rm R}) \propto {\rm i}\Lambda\epsilon$  to the first order of $\epsilon$. Although $\epsilon$ is a small parameter, the overall factor $\Lambda$ corresponding to the mass scale of right-handed neutrinos can be quite large, giving rise to a sizable value of ${\rm Det}(M^{}_{\rm R})$. On the other hand, Eq.~(\ref{eq:mnu0}) indicates that if $\widehat{g}^{}_3 = -\widehat{g}^{}_2/4$, only the (1,1)-element of $\widetilde{M}^{}_\nu$ survives. As can be seen from the top-right panel of Fig.~\ref{fig:figPara}, the ratio $\widehat{g}^{}_2/\widehat{g}^{}_3 = \widetilde{g}^{}_2/\widetilde{g}^{}_3$ is indeed around $-4$, corresponding to $\widetilde{g}^{}_3/\widetilde{g}^{}_2 = -0.25$ that is not far from the best-fit value $\widetilde{g}^{}_3/\widetilde{g}^{}_2 \approx -0.28$. Inspired by this observation, we further assume $\widehat{g}^{}_3 = -\widehat{g}^{}_2/4$ to hold for the stabilizer $\tau = {\rm i}$, and introduce perturbations to this identity when $\tau$ deviates from ${\rm i}$.

In the following discussions, we consider explicit perturbations to both $\tau = {\rm i}$ and $\widehat{g}^{}_3 = -\widehat{g}^{}_2/4$, and explore their implications for lepton masses, flavor mixing angles and CP-violating phases with some reasonable approximations.
\begin{itemize}
\item First, we introduce the perturbation to the stabilizer, i.e., $\tau = {\rm i}+ \epsilon$, where $\epsilon \equiv \epsilon^{}_{\rm R} + {\rm i}\epsilon^{}_{\rm I}$ is a complex parameter with $|\epsilon| = \sqrt{\epsilon^2_{\rm R} + \epsilon^2_{\rm I}} \ll 1$ and $\epsilon^{}_{\rm R}$ ($\epsilon^{}_{\rm I}$) being the real (imaginary) part. In the presence of this perturbation, the ratios of two adjacent basis vectors of ${\cal M}^{}_1[\Gamma(5)]$ read
\begin{eqnarray}
\frac{\widehat{e}^{}_2(\rm i+\epsilon)}{\widehat{e}^{}_1(\rm i+\epsilon)} = \frac{\widehat{e}^{}_3(\rm i+\epsilon)}{\widehat{e}^{}_2(\rm i+\epsilon)} =
\frac{\widehat{e}^{}_4(\rm i+\epsilon)}{\widehat{e}^{}_3(\rm i+\epsilon)} =
\frac{\widehat{e}^{}_5(\rm i+\epsilon)}{\widehat{e}^{}_4(\rm i+\epsilon)} =
\frac{\widehat{e}^{}_6(\rm i+\epsilon)}{\widehat{e}^{}_5(\rm i+\epsilon)} \approx A^{}_0 (1+1.245 \,{\rm i}\epsilon)\; ,
\label{eq:eratio_1}
% (5.6)
\end{eqnarray}
to the first order of $\epsilon$. With Eq.~(\ref{eq:eratio_1}), we can also write down the approximate expressions of all modular forms for $\tau = {\rm i} + \epsilon$. Then the neutrino mass matrix $\widetilde{M}^{}_\nu$ up to ${\cal O}(\epsilon^2_{})$ is
\begin{eqnarray}
\widetilde{M}^{}_\nu \approx -\frac{\widehat{g}^2_1 v^2_{\rm u}}{\Lambda \epsilon}
\left(\begin{matrix}
0.089\,{\rm i}-0.331\epsilon && 0 && 0 \\
0 && -0.823\,{\rm i}\widehat{g}^2_2\epsilon^2_{} && \widehat{g}^{2}_2\epsilon(1.258+5.731\,{\rm i}\epsilon) \\
0 && \widehat{g}^{2}_2\epsilon(1.258+5.731\,{\rm i}\epsilon) && 2.332\,{\rm i}\widehat{g}^2_2\epsilon^2_{} \\
\end{matrix}\right)^\ast_{}\; ,
\label{eq:mnu1}
% (5.7)
\end{eqnarray}
where the identity $\widehat{g}^{}_3=-\widehat{g}^{}_2/4$ is assumed. In the present paper, we focus on the case where $\epsilon^{}_{\rm R} < 0$ while $\epsilon^{}_{\rm I} > 0$. It is then evident that $H^{}_\nu \equiv  \widetilde{M}^{}_\nu \widetilde{M}^{\dag}_\nu$ can be diagonalized via $U^{\dag}_{23}H^{}_{\nu}U^{}_{23}={\rm Diag}\,\{m^{2}_{1,0},m^{2}_{2,0},m^{2}_{3,0}\}$ by the following unitary matrix
\begin{eqnarray}
U^{}_{23} = \left(
\begin{matrix}
1 && 0 && 0 \\
0 && 1 && 0 \\
0 && 0 && -{\rm i}\\
\end{matrix}\right) \cdot
\left(
\begin{matrix}
1 && 0 && 0 \\
0 && \cos\theta^{}_{23} && -\sin\theta^{}_{23} \\
0 && \sin\theta^{}_{23} && \cos\theta^{}_{23}
\end{matrix}\right) \; ,
%\cdot
%\left(
%\begin{matrix}
%e^{\rm i \varphi^{}_1}_{} && 0 && 0 \\
%0 && e^{\rm i (\varphi^{}_2-\pi/2)}_{} && 0 \\
%0 && 0 && e^{\rm i \varphi^{}_2}_{}
%\end{matrix}\right) \; ,
\label{eq:u23}
% (5.8)
\end{eqnarray}
where $\sin\theta^{}_{23} \approx \sqrt{2}/2 + 0.212\epsilon^{}_{\rm R}$. This simple formula implies that $\theta^{}_{23}$ is approximately $45^\circ_{}$ when ${\rm Re}\,\tau = \epsilon^{}_{\rm R}$ is small in magnitude, which is in good agreement with numerical results in the middle-left panel of Fig.~\ref{fig:figPara}. In addition, three eigenvalues of $H^{}_\nu$ can be expressed as
\begin{eqnarray}
m^{2}_{1,0} & \approx & (0.00793-0.0590\epsilon^{}_{\rm I}-0.0934\epsilon^2_{\rm R})\mu^{2}_0 \; , \nonumber \\
m^{2}_{2,0} & \approx & (1.583 + 3.971\epsilon^{}_{\rm R}-14.42\epsilon^{}_{\rm I}+35.91\epsilon^2_{\rm R})\widehat{g}^4_2\epsilon^{2}_{\rm R}\mu^{2}_0 \; ,\nonumber \\
m^{2}_{3,0} & \approx & (1.583 - 3.971\epsilon^{}_{\rm R}-14.42\epsilon^{}_{\rm I}+35.91\epsilon^2_{\rm R})\widehat{g}^4_2\epsilon^{2}_{\rm R}\mu^{2}_0 \; ,
\label{eq:ev1}
% (5.9)
\end{eqnarray}
where $\mu^{}_0 \equiv (\widehat{g}^{2}_1v^2_{\rm u})/(\Lambda |\epsilon|)$.  Note that  we have retained the terms up to next-to-leading order of $\epsilon^{}_{\rm R}$ but only the leading order terms of $\epsilon^{}_{\rm I}$. This treatment has been guided by the numerical results, which demonstrate that $|\epsilon^{}_{\rm R}|$ is about ten times larger than $\epsilon^{}_{\rm I}$ in the allowed parameter space. Eq.~(\ref{eq:ev1}) indicates that the values of $m^2_{2,0}$ and $m^2_{3,0}$ are proportional to $\widehat{g}^2_2 \epsilon^{2}_{\rm R}$, which should be highly suppressed because of a small value of $\epsilon^{}_{\rm R}$. However, the coefficients in the parentheses on the right-hand side of Eq.~(\ref{eq:ev1}) for $m^2_{2,0}$ and $m^2_{3,0}$ can be much larger than that for $m^{2}_{1,0}$. Therefore, it is possible to have the feasible parameter space of $\widehat{g}^{}_2$ where the normal mass ordering is allowed.

\item From previous discussions, we have observed that the perturbation to $\tau={\rm i}$ gives $\theta^{}_{23} \approx 45^{\circ}_{}$, and the identity $\widehat{g}^{}_3 = -\widehat{g}^{}_2/4$ leads to zero values of $\theta^{}_{12}$ and $\theta^{}_{13}$. In order to accommodate realistic mixing angles, we have to break the identity by assuming $\widehat{g}^{}_3 = -\widehat{g}^{}_2/4+\kappa$, where $\kappa$ is another perturbative parameter. To the first order of $\kappa$, the matrix $H^{}_\nu$ after the $(2,3)$-rotation is modified to be $H^{\prime}_{\nu} = \widehat{H}^{}_{\nu}+\Delta H^{}_\nu$ with $\widehat{H}^{}_{\nu} = {\rm Diag}\,\{m^2_{1,0},m^2_{2,0},m^2_{3,0}\}$, where the leading-order mass eigenvalues have been given in Eq.~(\ref{eq:ev1}), and the perturbation matrix $\Delta H^{}_\nu$ is given by
\begin{eqnarray}
\Delta H^{}_\nu \approx \mu^{2}_0 \kappa \left[0.00761\left(
\begin{matrix}
0 && -{\rm i} && +{\rm i} \\
+{\rm i} && 0 && 0 \\
-{\rm i} && 0 && 0 \\
\end{matrix}\right)+0.108\,\widehat{g}^2_{2}\left(
\begin{matrix}
0 && -{\rm i}\epsilon^*_{} && -{\rm i}\epsilon \\
+{\rm i}\epsilon^{}_{} && 0 && 0 \\
+{\rm i}\epsilon^\ast_{} && 0 && 0
\end{matrix}\right)\right]\; .
\label{eq:deltaHp}
% (5.10)
\end{eqnarray}
One can numerically check that the $(1,3)$-element of $\Delta H^{}_\nu$ is about ten times larger than its $(1,2)$-element if both $\epsilon$ and $\widehat{g}^{}_{2}$ are within their individual $3\sigma$ allowed range. Therefore, we implement a sequence of rotations, namely, the $(1,3)$-rotation followed by the $(1,2)$-rotation, on $H^\prime_\nu$ to diagonalize it. For the $(1,3)$-rotation, the unitary matrix $U^{}_{13}$ approximates to
\begin{eqnarray}
U^{}_{13} = \left(
\begin{matrix}
e^{{\rm i}\varphi^{}}_{} && 0 && 0 \\
0 && 1 && 0 \\
0 && 0 && 1 \\
\end{matrix}\right) \cdot
\left(
\begin{matrix}
\cos\theta^{}_{13} && 0 && -\sin\theta^{}_{13} \\
0 && 1 && 0 \\
\sin\theta^{}_{13} && 0 && \cos\theta^{}_{13}
\end{matrix}\right) \; ,
% (5.12)
\end{eqnarray}
where
\begin{eqnarray}
\sin\theta^{}_{13} \approx \frac{\mu^{2}_0 |\kappa|\left(0.00761-0.108\,\widehat{g}^2_2 \epsilon^{}_{\rm R}\right)}{m^2_{3,0}-m^2_{1,0}} \; , \quad \varphi \approx \arctan\left(\frac{0.0705-\widehat{g}^2_2 \epsilon^{}_{\rm R}}{\widehat{g}^2_2 \epsilon^{}_{\rm I}}\right) \; .
\label{eq:s13phi}
% (5.13)
\end{eqnarray}
From Eq.~(\ref{eq:s13phi}) we see that $\theta^{}_{13}$ is proportional to the new perturbative parameter $\kappa$, so it is always possible to get $\sin\theta^{}_{13} \sim 0.15$ by adjusting properly the value of $\kappa$. It is apparent that $\Delta H^{}_\nu$ also induces some corrections to the eigenvalues $m^2_{1,0}$ and $m^2_{3,0}$, which turn out to be proportional to $\kappa^2_{}$ and thus can be neglected. After the above $(1,3)$-rotation is applied to $H^\prime_{\nu}$, we have
\begin{eqnarray}
H^{\prime\prime}_\nu = U^\dag_{13}H^\prime_\nu U^{}_{13} \approx
\left(\begin{matrix}
m^2_{1,0} && -0.108 \, \mu^2_0  \widehat{g}^2_2 \epsilon^{}_{\rm I} \kappa e^{-{\rm i}\varphi}_{} && 0 \\
-0.108 \, \mu^2_0  \widehat{g}^2_2 \epsilon^{}_{\rm I} \kappa e^{{\rm i}\varphi}_{}  && m^2_{2,0} && 0 \\
0 && 0 && m^2_{3,0} \\
\end{matrix}\right)  \; ,
\label{eq:Hpp}
% (5.13)
\end{eqnarray}
where the approximation $\cos \theta^{}_{13} \approx 1$ has been made. Note that the modulus of the off-diagonal element $0.108 \,\mu^2_0 \widehat{g}^2_2 \epsilon^{}_{\rm I} \kappa$ is much smaller than $m^2_{1,0}$ and $m^2_{2,0}$ in magnitude, but the high degeneracy between $m^2_{1,0}$ and $m^2_{2,0}$ ensures a relatively large mixing angle $\theta^{}_{12}$. By using the degenerate perturbation theory, we obtain
\begin{eqnarray}
m^2_{1} & \approx & m^2_{1,0}-0.108\,\mu^2_0  \widehat{g}^2_2 \epsilon^{}_{\rm I} |\kappa| \; , \nonumber \\
m^2_{2} & \approx & m^2_{2,0} +0.108\,\mu^2_0  \widehat{g}^2_2 \epsilon^{}_{\rm I} |\kappa| \; ,\nonumber \\
m^2_{3} & \approx & m^2_{3,0} \; ,
\label{eq:ev2}
% (5.14)
\end{eqnarray}
which are the final results for three light neutrino masses. Meanwhile, the unitary matrix $U^{}_{12}$ to diagonalize $H^{\prime\prime}_{\nu}$ is found to be
%%%%%%%%%%%%%%%%%%%%%%%%%%%%%%% Fig. 2 %%%%%%%%%%%%%%%%%%%%%%%%%%%%%%%%
\begin{figure}[t!]
	\centering		\includegraphics[width=1\textwidth]{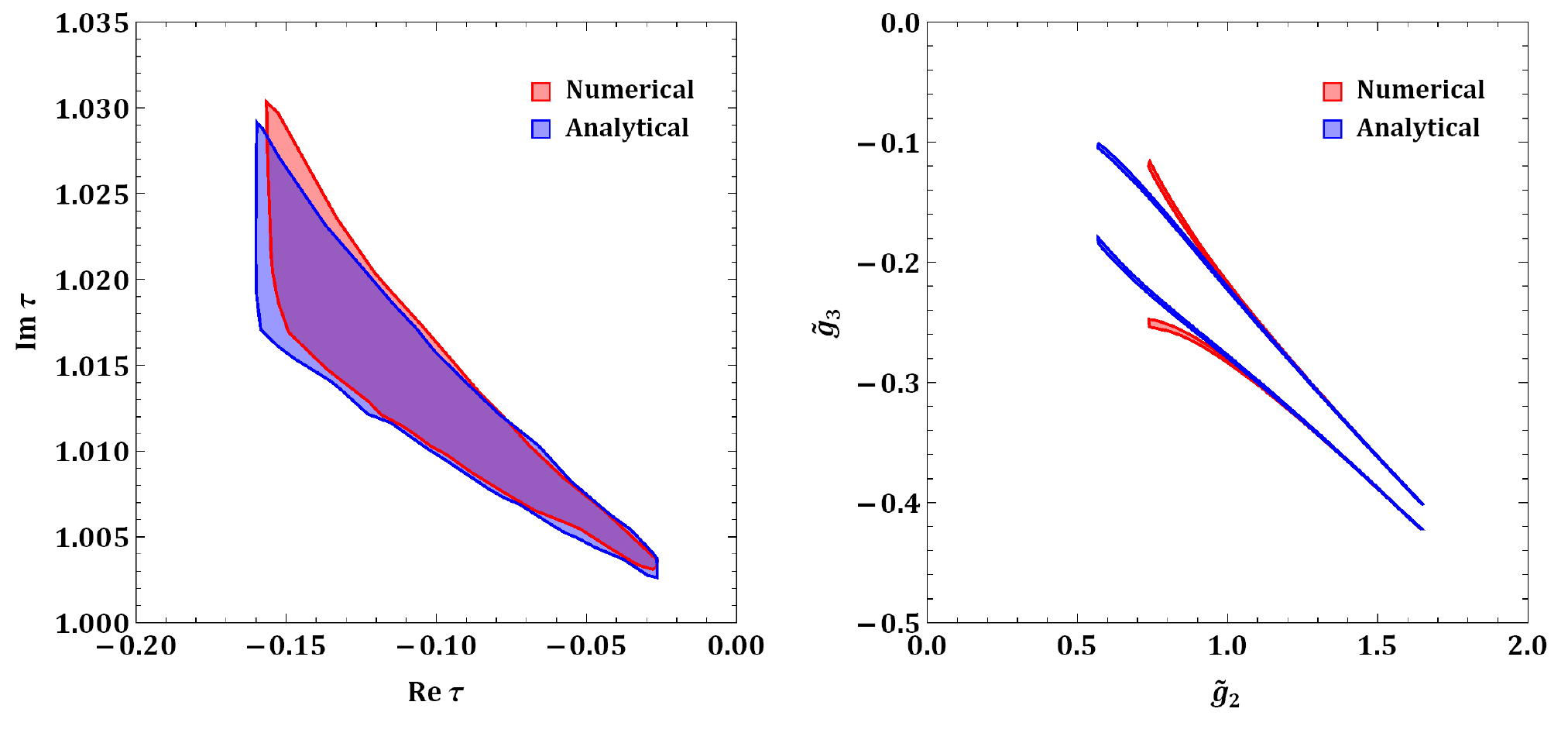}
	\vspace{-0.8cm}
	\caption{The $3\sigma$ allowed regions of $\{{\rm Re}\,\tau, {\rm Im}\,\tau\}$ and $\{\widetilde{g}^{}_{2},\widetilde{g}^{}_{3}\}$ obtained by inputting the approximate analytical formulas of three neutrino masses and mixing angles are represented by blue shaded areas. For comparison, the allowed regions obtained by exact numerical calculations are plotted in red.}
	\label{fig:ana_num} %% label for entire figure
\end{figure}
%%%%%%%%%%%%%%%%%%%%%%%%%%%%%%%%%%%%%%%%%%%%%%%%%%%%%%%%%%%%%%%%%%%%%%%
\begin{eqnarray}
U^{}_{12}=\left(
\begin{matrix}
e^{-{\rm i}\varphi} && 0 && 0 \\
0 && 1 && 0 \\
0 && 0 && 1 \\
\end{matrix}\right) \cdot
\left(
\begin{matrix}
\cos\theta^{}_{12} && -\sin\theta^{}_{12} && 0 \\
\sin\theta^{}_{12} && \cos\theta^{}_{12} && 0 \\
0 && 0 && 1
\end{matrix}\right) \; ,
\label{eq:U12}
% (5.15)
\end{eqnarray}
with
\begin{eqnarray}
\sin^2_{}\theta^{}_{12} \approx \dfrac{0.108\, \mu^2_0  \widehat{g}^2_2 \epsilon^{}_{\rm I} |\kappa|}{m^{2}_{2,0}-m^{2}_{1,0}+0.216\,\mu^2_0  \widehat{g}^2_2 \epsilon^{}_{\rm I} |\kappa|} \; ,
\label{eq:s12}
% (5.16)
\end{eqnarray}
implying that the value of $\sin^2_{}\theta^{}_{12}$ can be greatly enhanced if $m^{2}_{2,0}$ is very close to $m^{2}_{1,0}$. For example, if $m^{2}_{2,0}-m^{2}_{1,0} \approx 0.108\, \mu^2_0  \widehat{g}^2_2 \epsilon^{}_{\rm I} |\kappa|$ or equivalently $\Delta m^2_{21} \approx 0.324\, \mu^2_0  \widehat{g}^2_2 \epsilon^{}_{\rm I} |\kappa|$, we will obtain $\sin^2_{}\theta^{}_{12} \approx 1/3$, which agrees well with the experimental observation. To illustrate how well the analytical approximations are in agreement with exact numerical results, we have also scanned over the model parameters by using the approximate expressions of three neutrino masses and mixing angles. The $3\sigma$ allowed regions of $\{{\rm Re}\,\tau,{\rm Im}\,\tau,\widetilde{g}^{}_2, \widetilde{g}^{}_3\}$ are shown in Fig.~\ref{fig:ana_num}, where an excellent agreement with the exact numerical results can be found for $|{\rm Re}\,\tau| \lesssim 0.1$ and $\widetilde{g}^{}_{2} \gtrsim 1.0$.
\end{itemize}

%%%%%%%%%%%%%%%%%%%%%%%%%%%%%%% Fig. 3 %%%%%%%%%%%%%%%%%%%%%%%%%%%%%%%%
\begin{figure}[t!]
	\centering		\includegraphics[width=0.32\textwidth]{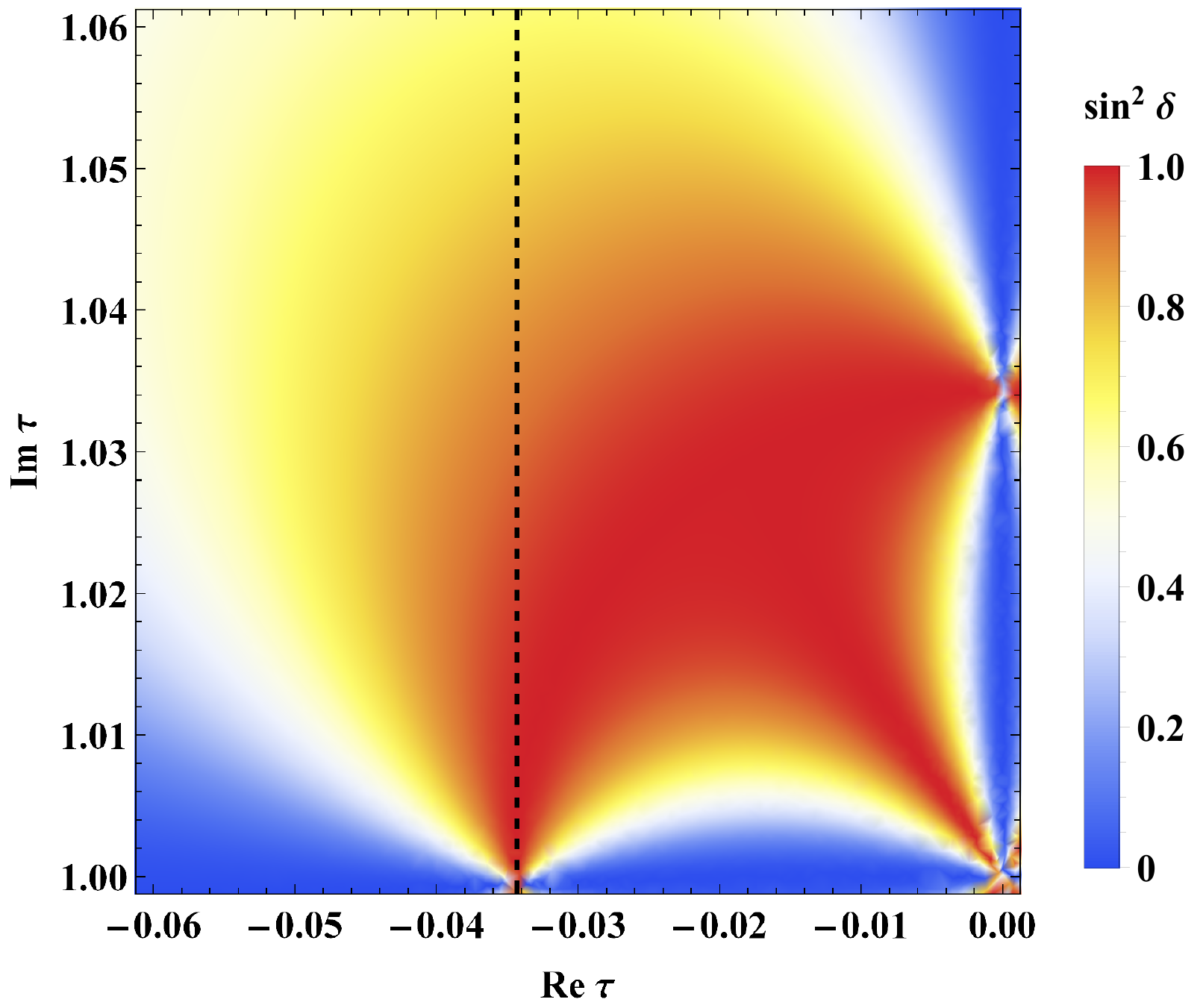}
	\includegraphics[width=0.32\textwidth]{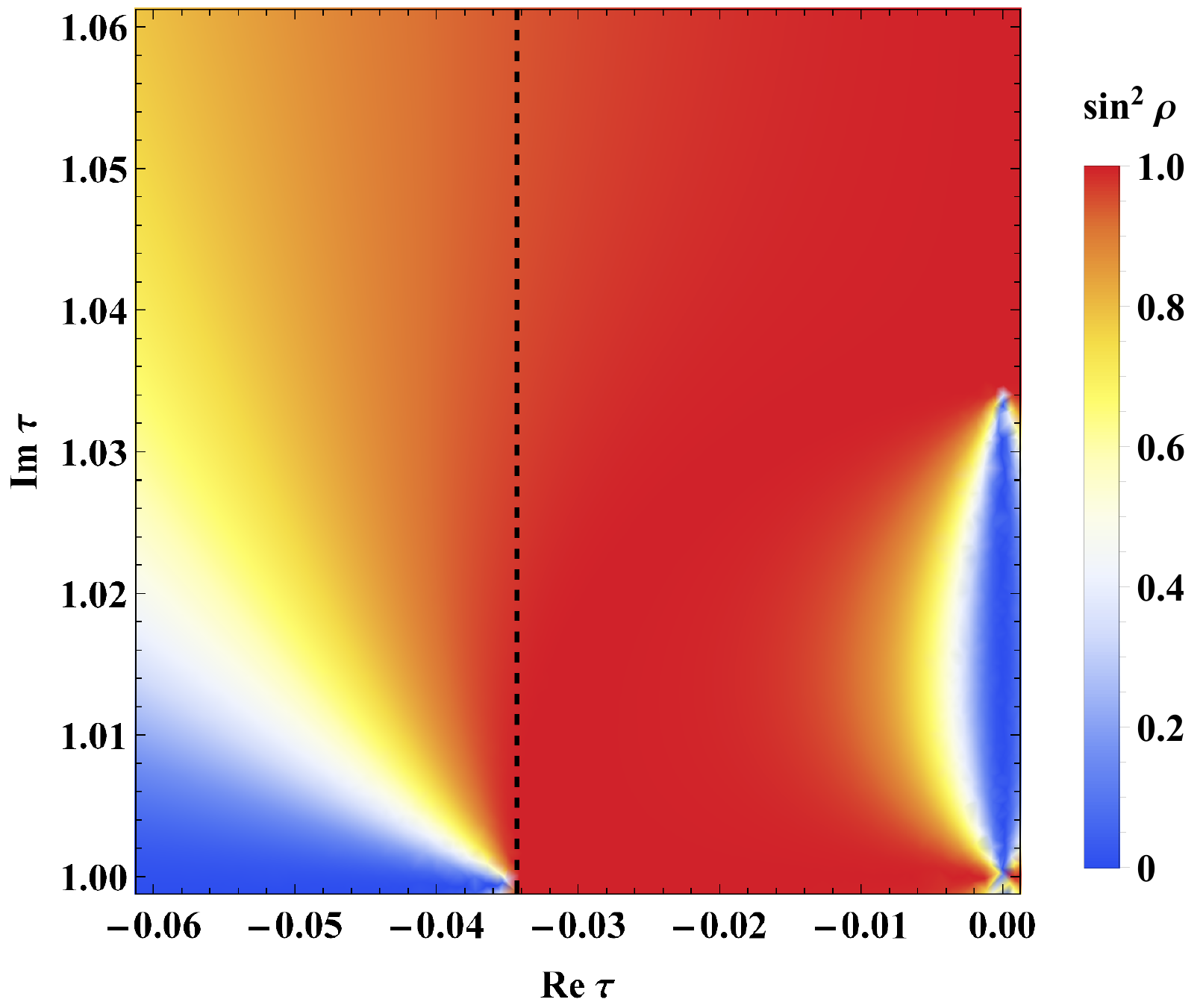}
	\includegraphics[width=0.32\textwidth]{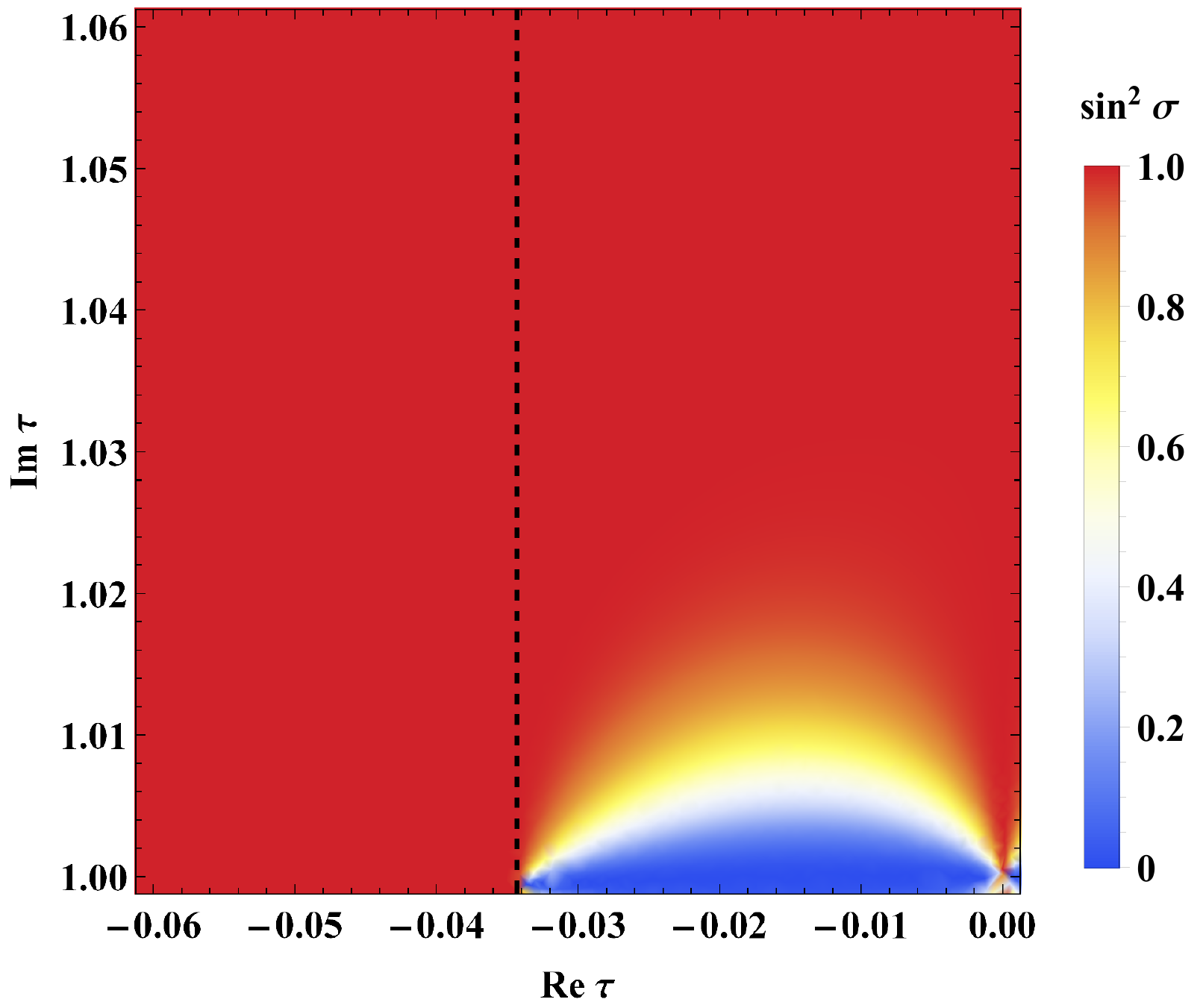}
	\vspace{0.2cm}
	\caption{The distributions of $\{\sin^2_{}\delta, \sin^2_{}\rho, \sin^2_{}\sigma\}$ in the vicinity of $\tau = {\rm i}$, where $\widetilde{g}^{}_2 = 1.456$ and $\widetilde{g}^{}_3 = -0.350$ are fixed. The vertical dotted line corresponds to ${\rm Re}\,\tau = -0.0343$, which is the typical value of ${\rm Re}\,\tau$ in its $3\sigma$ allowed parameter space.}
	\label{fig:cp} %% label for entire figure
\end{figure}
%%%%%%%%%%%%%%%%%%%%%%%%%%%%%%%%%%%%%%%%%%%%%%%%%%%%%%%%%%%%%%%%%%%%%%%

Combining three unitary matrices $U^{}_{23}$, $U^{}_{13}$ and $U^{}_{12}$ together, we finally obtain the Pontecorvo-Maki-Nakagawa-Sakata (PMNS) matrix~\cite{Pontecorvo:1957cp,Maki:1962mu}
\begin{eqnarray}
U=U^{}_{23}U^{}_{13}U^{}_{12} = \left(
\begin{matrix}
c^{}_{12}c^{}_{13} && s^{}_{12}c^{}_{13} && s^{}_{13}e^{{\rm i}\varphi}_{} \\
-s^{}_{12}c^{}_{23}-c^{}_{12}s^{}_{23}s^{}_{13}e^{-{\rm i}\varphi} && c^{}_{12}c^{}_{23}-s^{}_{12}s^{}_{23}s^{}_{13}e^{-{\rm i}\varphi}_{} && s^{}_{23}c^{}_{13} \\
s^{}_{12}s^{}_{23}-c^{}_{12}c^{}_{23}s^{}_{13}e^{-{\rm i}\varphi} && -c^{}_{12}s^{}_{23}-s^{}_{12}c^{}_{23}s^{}_{13}e^{-{\rm i}\varphi} &&
c^{}_{23}c^{}_{13}
\end{matrix}\right) \; ,
\label{eq:UPMNS}
% (5.17)
\end{eqnarray}
where $s^{}_{ij} \equiv \sin \theta^{}_{ij}$ and $c^{}_{ij} \equiv \cos \theta^{}_{ij}$ (for $ij = 12, 13, 23$) have been defined, and the unphysical phases have been eliminated by redefining the phases of the charged-lepton fields. Comparing Eq.~(\ref{eq:UPMNS}) with the standard parametrization of $U$~\cite{PDG2020}, we can extract the Dirac CP-violating phase $\delta$ as
\begin{eqnarray}
\delta = 2\pi - \varphi \approx 2\pi - \arctan\left(\frac{0.0705-\widehat{g}^2_2 \epsilon^{}_{\rm R}}{\widehat{g}^2_2 \epsilon^{}_{\rm I}}\right) \; .
\label{eq:delta}
% (5.18)
\end{eqnarray}
Since the value of $|\epsilon^{}_{\rm R}|$ is much larger than $\epsilon^{}_{\rm I}$ in the allowed parameter space, we obtain a nearly-maximal Dirac CP-violating phase, i.e., $\delta \approx 270^\circ_{}$. Two Majorana CP-violating phases can be figured out by diagonalizing the neutrino mass matrix via $U^\dagger \widetilde{M}^{}_\nu U^* = {\rm Diag}\{m^{}_1 e^{{\rm i}\rho}, m^{}_2 e^{{\rm i}\sigma}, m^{}_3\}$. It turns out that
\begin{eqnarray}
\rho \approx \frac{\pi}{2}+\arctan\left(\frac{\epsilon^{}_{\rm I}}{\epsilon^{}_{\rm R}}\right) \; , \quad \sigma \approx \frac{\pi}{2}-\arctan\left(\frac{\epsilon^{}_{\rm I}}{\epsilon^{}_{\rm R}}\right) \; ,
% (5.19)
\label{eq:Majorana}
\end{eqnarray}
 which are both close to $90^\circ_{}$ since $\epsilon^{}_{\rm I}\ll |\epsilon^{}_{\rm R}|$. In the left panel of Fig.~\ref{fig:cp}, we present the distribution of $\sin^2_{}\delta$ in the vicinity of $\tau = {\rm i}$ where $\widetilde{g}^{}_2 = 1.456$ and $\widetilde{g}^{}_3 = -0.350$ are fixed for illustration. For the chosen values of $\widetilde{g}^{}_2$ and $\widetilde{g}^{}_3$, the $3\sigma$ allowed value of ${\rm Re}\,\tau$ is tightly restricted to be around $-0.0343$, which is denoted by the vertical dotted line in Fig.~\ref{fig:cp}. Along this dotted line, the value of $\sin^2_{}\delta$ goes down as ${\rm Im}\,\tau$ increases, which can be well described by Eq.~(\ref{eq:delta}). For completeness, we also show the distributions of $\sin^2_{}\rho$ and $\sin^2_{}\sigma$ around $\tau = {\rm i}$ in the right two panels of Fig.~\ref{fig:cp}. Note that the values of $\sin^2_{}\rho$ and $\sin^2_{}\sigma$ along some parts of the imaginary axis as well as the lower boundary of the fundamental domain ${\cal G}$ can be maximal, corresponding to $\rho = \sigma =90^\circ_{}$. However, CP violation associated with Majorana phases should be always proportional to $\sin 2\rho$ or $\sin 2\sigma$, implying CP conservation along the imaginary axis and the lower boundary of ${\cal G}$.

 %%%%%%%%%%%%%%%%%%%%%%%%%%%%%%% Fig. 4 %%%%%%%%%%%%%%%%%%%%%%%%%%%%%%%%
 \begin{figure}[t!]
 	\centering		\includegraphics[width=0.58\textwidth]{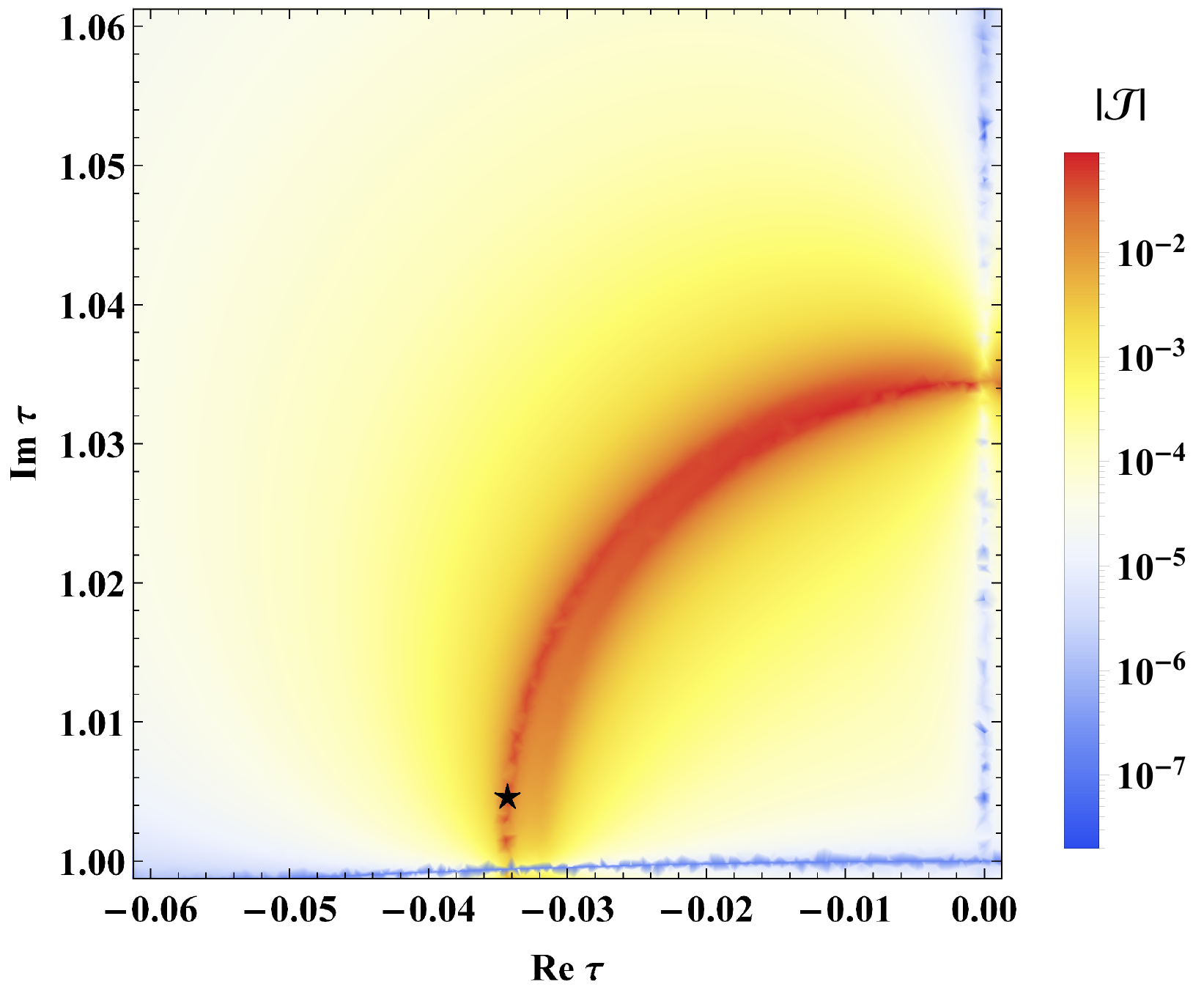}
 	\vspace{0.2cm}
 	\caption{The distribution of the absolute value of Jarlskog invariant $|{\cal J}|$ in the vicinity of $\tau =  {\rm i}$, where $\widetilde{g}^{}_2 = 1.456$ and $\widetilde{g}^{}_3 = -0.350$ are fixed. The black star corresponds to ${\rm Re}\,\tau = -0.0343$ and ${\rm Im}\,\tau = 1.00466$, which are the typical values of $\{{\rm Re}\,\tau,{\rm Im}\,\tau\}$ in their $3\sigma$ allowed parameter space.}
 	\label{fig:jar} %% label for entire figure
 \end{figure}
 %%%%%%%%%%%%%%%%%%%%%%%%%%%%%%%%%%%%%%%%%%%%%%%%%%%%%%%%%%%%%%%%%%%%%%%

One can also adopt a parametrization-independent approach to measure the magnitude of CP violation. For instance, the Jarlskog invariant ${\cal J} \equiv {\rm Im}\left[U^{}_{e 1} U_{e 2}^{*} U_{\mu 1}^{*} U_{\mu 2}^{}\right]$ for CP violation in leptonic sector~\cite{Jarlskog:1985ht,Wu:1985ea} can be calculated via
\begin{eqnarray}
\mathcal{I}_{1}=-6 \mathrm{i} \Delta m^{2}_{21} \Delta m^{2}_{31} \Delta m^{2}_{32} \Delta m^{2}_{e \mu} \Delta m^{2}_{\mu \tau} \Delta m^{2}_{\tau e} \mathcal{J} \; ,
\label{eq:jar}
% (5.19)
\end{eqnarray}
where $\Delta m^{2}_{\alpha \beta} \equiv m_{\alpha}^{2}-m_{\beta}^{2}$ (for $\alpha, \beta=e,\mu,\tau$) denote the mass-squared differences of charged leptons and $\mathcal{I}_{1} \equiv \operatorname{Tr}\left\{\left[H_{\nu}, H_{l}\right]^{3}\right\}$ is one of the CP-odd weak-basis invariants~\cite{Bernabeu:1986fc, Branco:1986gr, Yu:2019ihs,Yu:2020gre}. Substituting the approximate expressions of $H^{}_\nu$ and $H^{}_l$ into ${\cal I}^{}_1$ and using Eq.~(\ref{eq:jar}), we obtain
\begin{eqnarray}
{\cal J} \approx \frac{0.00963\,\mu^{6}_{0}\kappa^2_{} \widehat{g}^6_{2}\epsilon^3_{\rm R}(\epsilon^{}_{\rm I}-1.474\,\epsilon^{2}_{\rm R})}{\Delta m^2_{21}\Delta m^2_{31}\Delta m^2_{32}} \; ,
\label{eq:jar1}
% (5.20)
\end{eqnarray}
where one can observe that ${\cal J}$ may be highly suppressed due to a tiny numerator on the right-hand side. Therefore, a sizable value of ${\cal J}$ can be achieved if $\Delta m^2_{21}$ is small enough. For example, if we take $\Delta m^2_{21} = 0.324\, \mu^2_0  \widehat{g}^2_2 \epsilon^{}_{\rm I} |\kappa|$, corresponding to $\sin^2_{}\theta^{}_{12} = 1/3$ as discussed below Eq.~(\ref{eq:s12}), then Eq.~(\ref{eq:jar1}) can be simplified to
\begin{eqnarray}
{\cal J} \approx \dfrac{0.283\,r^2_{} \epsilon^3_{\rm R}(\epsilon^{}_{\rm I}-1.474\,\epsilon^2_{\rm R})}{|\kappa|\epsilon^3_{\rm I}(1-r)} \; ,
\label{eq:jar2}
% (5.21)
\end{eqnarray}
where the definition $ r \equiv \Delta m^2_{21}/\Delta m^2_{31}$ in the NO case has been used. The maximal CP violation with $\delta \approx270^\circ_{}$ corresponds to ${\cal J} =s^{}_{12} c^{}_{12} s^{}_{23} c^{}_{23} s^{}_{13} c_{13}^{2} \sin \delta \approx -0.03$, if all the mixing angles take their individual best-fit values from the global-fit analysis as shown in Table~\ref{table:gfit}. Assuming $\epsilon^{}_{\rm R} \approx -0.03$ and $\epsilon^{}_{\rm I} \approx 0.003$ in Eq.~(\ref{eq:jar2}), we can find that ${\cal J} \approx -0.03$ requires $\kappa \approx 15r^2 \approx 0.015$. The distribution of $|{\cal J}|$ in the vicinity of  $\tau = {\rm i}$ with  $\widetilde{g}^{}_2 = 1.456$ and $\widetilde{g}^{}_3 = -0.350$ is shown in Fig.~\ref{fig:jar}, where one can clearly see that the red ring of radius $\sqrt{\epsilon^2_{\rm R}+\epsilon^2_{\rm I}} \sim 0.035$ corresponds to the relatively large values of ${\cal J}$. The $3\sigma$ allowed values of $\{{\rm Re}\,\tau,{\rm Im}\,\tau\}$ denoted by the black star are exactly located at this ring, indicating the nearly-maximal CP violation.

Before closing this section, let us briefly discuss why the IO case is not permitted in our model. For the IO case, we can also introduce explicit perturbations to $\tau = {\rm i}$ in a similar way. After the first $(2,3)$-rotation, the $(1,2)$-element of $H^\prime_\nu$ becomes much larger than the counterpart in the NO case. As a consequence, if we require $\sin^2_{} \theta^{}_{12} \sim 1/3$, $r \equiv \Delta m^2_{21}/|\Delta m^{2}_{32}|$ will be close to one, which is apparently incompatible with the experimental observation.

\section{Radiative Corrections}\label{sec:RGE}

The modular symmetry and seesaw mechanism are supposed to work at a very high energy scale $\Lambda^{}_{\rm SS}$, whereas the leptonic flavor mixing parameters are measured in the neutrino oscillation experiments whose energy scales are much lower than the electroweak scale characterized by the mass of the $Z$ gauge boson $m^{}_{Z} \approx 91.2~{\rm GeV}$, varying from several MeV to GeV. In order to confront theoretical predictions with experimental observations, one has to take into account the radiative corrections to physical parameters via their RG equations. It is well known that the RG running effects can be significant, especially for large values of $\tan\beta$ and (or) nearly-degenerate neutrino masses~\cite{Ohlsson:2013xva}. Similar conclusions have also been reached in the modular-invariant flavor models~\cite{Wang:2020dbp,Criado:2018thu}. As indicated in Eq.~(\ref{eq:s12}), the value of $\sin^2_{}\theta^{}_{12}$ in our model is very sensitive to small changes of model parameters, particularly in the region close to $\tau = {\rm i}$. Hence it is reasonable to expect large corrections from RG running even if the value of $\tan\beta$ is relatively small.
%%%%%%%%%%%%%%%%%%%%%%%%%%%%%% Fig. 4 %%%%%%%%%%%%%%%%%%%%%%%%%%%%%%%%
\begin{figure}[t!]
	\centering		\includegraphics[width=0.96\textwidth]{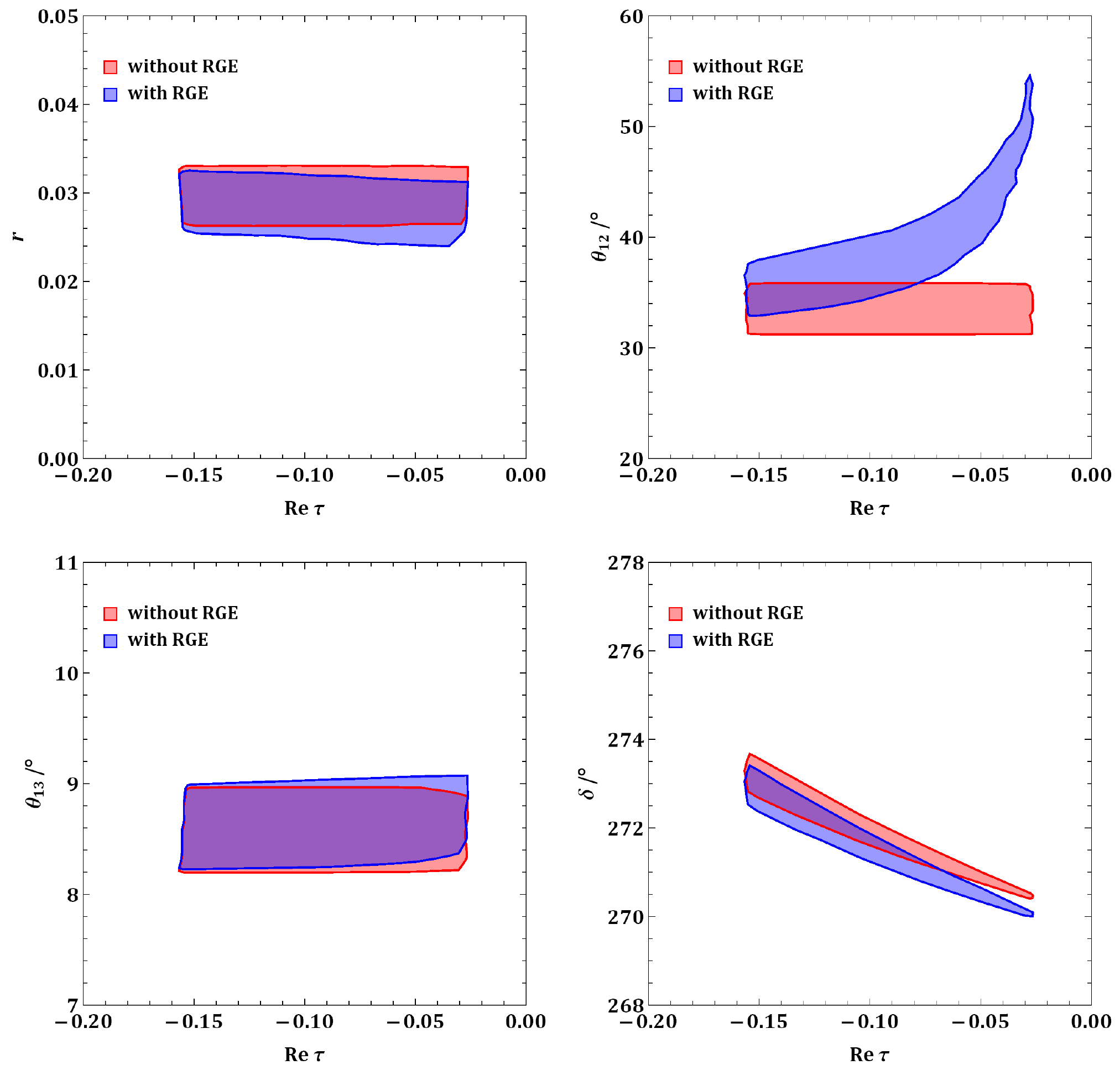}
	\vspace{-0.5cm}
	\caption{Correlations between low-energy observables $\{r,\theta^{}_{12},\theta^{}_{13},\delta\}$ and the model parameter ${\rm Re}\,\tau$, where the red-shaded areas are gained by inputting the $3\sigma$ allowed parameter space in Sec.~\ref{sec:num} without the RG running effects, while the blue-shaded areas are obtained with same input but by taking account of the RG running effects from the GUT scale to the electroweak scale.}
	\label{fig:RGE} %% label for entire figure
\end{figure}
%%%%%%%%%%%%%%%%%%%%%%%%%%%%%%%%%%%%%%%%%%%%%%%%%%%%%%%%%%%%%%%%%%%%%%%

For definiteness, we first carry out a quantitative analysis of RG running effects on the mixing parameters from the GUT scale $\Lambda^{}_{\rm SS} \sim \Lambda^{}_{\rm GUT}$ to the electroweak scale, where $\tan\beta = 10$ will be assumed. The RG running effects from the electroweak scale to the energy scale of oscillation experiments are not significant and thus can be neglected. In the basis where the charged-lepton Yukawa coupling matrix $\widetilde{Y}^{}_l \equiv {\rm Diag}\{y^{}_e, y^{}_\mu, y^{}_\tau\}$ is diagonal with $y^{}_\alpha = \sqrt{2} m^{}_\alpha/v^{}_{\rm d}$ (for $\alpha = e, \mu, \tau$), the Dirac neutrino Yukawa coupling matrix becomes $\widetilde{Y}^{}_\nu = U^\dagger_l Y^{}_\nu$ with $Y^{}_\nu = \sqrt{2} M^{}_{\rm D}/v^{}_{\rm u}$. After integrating out heavy Majorana neutrinos below the scale $\Lambda^{}_{\rm SS}$, we obtain the effective neutrino mass parameter ${\cal M} \equiv - U^{\dag}_{l}Y^{}_{\nu}M^{-1}_{\rm R}Y^{\rm T}_{\nu}U^\ast_l$. In the minimal supersymmetric SM, the radiative corrections to $\widetilde{Y}^{}_l$ and ${\cal M}$ can be described by the one-loop RG equations~\cite{Machacek:1983fi, Arason:1991ic, Castano:1993ri, Chankowski:1993tx, Babu:1993qv, Antusch:2001ck}
\begin{eqnarray}
16\pi^2_{}\frac{{\rm d} \widetilde{Y}^{}_l }{{\rm d} t} &=&  \left[\alpha^{}_{l} + 3 \left(\widetilde{Y}^{}_l \widetilde{Y}^{\dag}_{l}\right)\right]\widetilde{Y}^{}_l \; , \label{eq:MRGE1}\\
16\pi^2_{}\frac{{\rm d} {\cal M}}{{\rm d} t} &=& \alpha^{}_{\nu}{\cal M} + \left[ \left(\widetilde{Y}^{}_l \widetilde{Y}^{\dag}_{l}\right){\cal M}+{\cal M} \left(\widetilde{Y}^{}_l \widetilde{Y}^{\dag}_{l}\right)^{\rm T}_{}\right] \; ,
\label{eq:MRGE}
% (6.1)-(6.2)
\end{eqnarray}
where $t \equiv \ln (\mu/\Lambda^{}_{\rm SS})$ with $\mu$ being the renormalization scale, $\alpha^{}_{l} = - 1.8 \mathfrak{g}^2_1 - 3 \mathfrak{g}^2_2 + 3 {\rm Tr}(Y^{}_{\rm d}Y^{\dag}_{\rm d}) + {\rm Tr}(\widetilde{Y}^{}_l \widetilde{Y}^{\dag}_l)$ and $\alpha^{}_{\nu}= -1.2 \mathfrak{g}^2_1 - 6 \mathfrak{g}^2_2 + 6 {\rm Tr}(Y^{}_{\rm u}Y^{\dag}_{\rm u})$. Notice that $\mathfrak{g}^{}_{2}$ and $\mathfrak{g}^{}_{1}$ represent respectively the ${\rm SU}(2)^{}_{\rm L}$ and ${\rm U}(1)^{}_{\rm Y}$ gauge couplings, $Y^{}_{\rm u}$ and $Y^{}_{\rm d}$ denote the up- and down-type quark Yukawa coupling matrices. In the chosen flavor basis where $\widetilde{Y}^{}_l$ is diagonal, only the diagonal elements on both sides of Eq.~(\ref{eq:MRGE1}) survive. Therefore, the radiative corrections to neutrino masses and lepton flavor mixing parameters are determined by the RG equation of ${\cal M}$~\cite{Mei:2003gn}. The solution to ${\cal M}$ in Eq.~(\ref{eq:MRGE}) can be formally written as
\begin{eqnarray}
{\cal M}(m^{}_Z) = I^{}_\nu \left(
\begin{matrix}
I^{}_e &&& 0 &&& 0 \\ 0 &&& I^{}_\mu &&& 0 \\ 0 &&& 0 &&& I^{}_\tau
\end{matrix}
\right) {\cal M}(\Lambda^{}_{\rm SS}) \left(
\begin{matrix}
I^{}_e &&& 0 &&& 0 \\ 0 &&& I^{}_\mu &&& 0 \\ 0 &&& 0 &&& I^{}_\tau
\end{matrix}
\right) \; ,
\label{eq:solRGE}
% (6.3)
\end{eqnarray}
where $I^{}_\nu$ and $I^{}_\alpha$ (for $\alpha = e, \mu, \tau$) are the evolution functions~\cite{Mei:2003gn}
\begin{eqnarray}
I^{}_\nu & = & {\rm exp}\left[-\frac{1}{16\pi^2_{}}\int^{\ln (\Lambda^{}_{\rm SS}/m^{}_Z)}_{0}\alpha^{}_{\nu}(t) {\rm d}t\right] \; , \nonumber \\
I^{}_\alpha & = & {\rm exp}\left[-\frac{1}{16\pi^2_{}}\int^{\ln (\Lambda^{}_{\rm SS}/m^{}_Z)}_{0} y^2_\alpha (t) {\rm d}t\right] \; .
\label{eq:evofun}
% (6.4)
\end{eqnarray}
While $I^{}_\nu$ in Eq.~(\ref{eq:solRGE}) affects only the absolute scale of light neutrino masses, $I^{}_\alpha$ could modify both neutrino masses and flavor mixing parameters. By using Eq.~(\ref{eq:solRGE}), one can establish the direct connection between the effective neutrino mass parameter at the high-energy scale $\Lambda^{}_{\rm SS}$ and that at the electroweak scale $m^{}_Z$.

To illustrate radiative corrections to flavor mixing parameters, we numerically solve the RG equations and search for the allowed parameter space by re-scanning over the model parameters. It turns out that the $3\sigma$ allowed regions of $\{{\rm Re}\,\tau,{\rm Im}\,\tau,\widetilde{g}^{}_2,\widetilde{g}^{}_3 \}$ obtained with or without radiative corrections are hardly distinguishable, implying that the RG running effects will almost not change the allowed parameter space. However, they can greatly modify the predictions for flavor mixing parameters if we fix the values of $\{{\rm Re}\,\tau,{\rm Im}\,\tau,\widetilde{g}^{}_2,\widetilde{g}^{}_3 \}$. To make this point clearer, we identify the $3\sigma$ allowed parameter space obtained without considering the RG running effects in Sec.~\ref{sec:num} as the input at the GUT scale, and calculate the predictions for neutrino masses and mixing parameters at the electroweak scale with radiative corrections. The correlations between low-energy observables $\{r,\theta^{}_{12},\theta^{}_{13},\delta\}$ and the model parameter ${\rm Re}\,\tau$ are shown in Fig.~\ref{fig:RGE}, where both the red- and blue-shaded areas are gained by inputting the $3\sigma$ allowed parameter space in Sec.~\ref{sec:num}, while only the blue-shaded areas correspond to the results with the radiative corrections. Notice that the correction to $\theta^{}_{23}$ is negligibly small, so we do not show the result of $\theta^{}_{23}$. As one can see from Fig.~\ref{fig:RGE}, the RG running effects can modify the predicted values of oscillation parameters to a large extent especially when $|{\rm Re}\,\tau|$ is small. For instance, if $|{\rm Re}\,\tau|$ is as small as $0.03$, the mixing angle $\theta^{}_{12}$ can reach $55^\circ_{}$, which is about $60\%$ larger than that without the RG running effects. 

To better understand why radiative corrections to $\theta^{}_{12}$ are so large, we consider the RG running effects on the model parameters. Since the electron and muon Yukawa couplings are extremely small, $I^{}_e \approx I^{}_\mu \approx 1$ holds as an excellent approximation and $I^{}_\tau$ contributes dominantly to the running effects on ${\cal M}$ in Eq.~(\ref{eq:solRGE}). With the help of Eq.~(\ref{eq:solRGE}), we can derive the approximate formula of $\widetilde{M}^{}_\nu (m^{}_Z)$ around the stabilizer $\tau = {\rm i}$ after including the RG running. Then, three light neutrino masses at the electroweak scale are
\begin{eqnarray}
m^2_{1}(m^{}_Z) & \approx & m^2_{1,0}-0.108\,I^{}_\tau\mu^2_0  \widehat{g}^2_2 \epsilon^{}_{\rm I} |\kappa| \; , \nonumber \\
m^2_{2}(m^{}_Z) & \approx & I^2_\tau m^2_{2,0} +0.108\,I^{}_\tau\mu^2_0  \widehat{g}^2_2 \epsilon^{}_{\rm I} |\kappa| \; ,\nonumber \\
m^2_{3}(m^{}_Z) & \approx & I^2_\tau  m^2_{3,0} \; ,
\label{eq:ev2RGE}
% (6.5)
\end{eqnarray}
and the expression of $\sin^2_{}\theta^{}_{12}(m^{}_Z)$ is approximately given by
\begin{eqnarray}
\sin^2_{}\theta^{}_{12}(m^{}_Z) \approx \dfrac{0.108\, I^{}_\tau \mu^2_0  \widehat{g}^2_2 \epsilon^{}_{\rm I} |\kappa|}{I^2_{\tau} m^{2}_{2,0}-m^{2}_{1,0}+0.216\,I^{}_\tau\mu^2_0  \widehat{g}^2_2 \epsilon^{}_{\rm I} |\kappa|} \; ,
\label{eq:s12RGE}
% (6.6)
\end{eqnarray}
which will be reduced to Eq.~(\ref{eq:s12}) for $I^{}_\tau = 1$. As pointed out in the last section, $m^2_{2,0}$ can be very close to $m^2_{1,0}$ when ${\rm Re}\,\tau$ approaches zero. Therefore, although $I^{}_\tau - 1 \approx -0.0015$ is extremely small in our case, $I^2_\tau m^2_{2,0} - m^2_{1,0}$ in the	denominator on the right-hand side of Eq.~(\ref{eq:s12RGE}) leads to large corrections to $\sin^2_{}\theta^{}_{12}$. For instance, let us set the values of model parameters to be $\epsilon = -0.0301 + 0.00359\,{\rm i}$, $\widehat{g}^{}_2 = 1.555$ and $\kappa=0.0125$, then Eq.~(\ref{eq:s12}) gives $\theta^{}_{12} \approx 31.5^\circ_{}$. On the other hand, if the radiative corrections are taken into consideration, Eq.~(\ref{eq:s12RGE}) will give $\theta^{}_{12}(m^{}_Z) \approx 50.1^\circ_{}$, where one can observe a remarkable modification of $\theta^{}_{12}$ due to the RG running effects. In conclusion, despite the fact that the RG running effects on the model parameters are inconspicuous, the low-energy observables are sensitive to small perturbations to free parameters in the vicinity of $\tau = {\rm i}$ and radiative corrections are actually playing a significant role.

Although our analytical calculations are performed in a specific model, they are indeed helpful in understanding sizable radiative corrections to the mixing angle $\theta^{}_{12}$ for nearly-degenerate neutrino masses in the most general case. The key point is that the large mixing angle $\theta^{}_{12}$ is essentially determined by the ratio of two small model parameters, to which radiative corrections themselves are tiny but change considerably the ratio.

\section{Summary}\label{sec:sum}

Modular flavor symmetries provide us with an attractive way to account for lepton flavor mixing and CP violation. Instead of just fitting the modulus $\tau$ to the experimental data, one can also start with one of some special values $\tau$ (i.e., the fixed points or stabilizers), at which residual symmetries are retained after the breaking of the global modular symmetry. Any realistic models for lepton masses, flavor mixing and CP violation require the modulus $\tau$ to slightly deviate from the stabilizers. Therefore, it is interesting to explore the basic properties of modular-symmetry models with a modulus $\tau$ in the vicinity of the stabilizer. Such an exploration is useful for understanding the phenomenological implications of the modular-symmetry models, and gives a clue to possible residual symmetries and the modular symmetry itself.

In this paper, we construct a feasible lepton flavor model based on the modular $A^\prime_5$ group combined with the gCP symmetry, and focus on the fixed point $\tau = {\rm i}$ where CP symmetry is preserved. A seemingly puzzling feature is that it predicts a nearly-maximal CP-violating phase in the region where $\tau = {\rm i} + \epsilon$ with $\epsilon$ being a small parameter. By treating $\epsilon$ as a perturbation to the stabilizer $\tau = {\rm i}$, we derive the approximate analytical expressions of light neutrino masses, flavor mixing angles and the CP-violating phases. We find that $\sin^2_{}\theta^{}_{12}$ depends on the mass-squared difference $\Delta m^2_{21}$, and the enhancement of $\sin^2_{}\theta^{}_{12}$ can be attributed to the high degeneracy between two neutrino masses $m^{}_1$ and $m^{}_2$. The approximate expression of $\tan\delta$ is found to be $- (0.0705-\widehat{g}^2_2 \epsilon^{}_{\rm R})/(\widehat{g}^2_2 \epsilon^{}_{\rm I})$, where $\epsilon^{}_{\rm R} \equiv {\rm Re}\,\epsilon$ is about ten times larger than $\epsilon^{}_{\rm I} \equiv {\rm Im}\,\epsilon$ in the $3\sigma$ allowed parameter space. This analytical result shows that a nearly-maximal CP-violating phase can be obtained near $\tau = {\rm i}$. We find that $\epsilon \approx -0.03 +  0.003\,{\rm i}$ is sufficient to give ${\cal J} \approx -0.03$, corresponding to the nearly-maximal CP-violating phase $\delta \approx 270^\circ$  when all the mixing angles are consistent with their observed values.

Since $\sin^2_{}\theta^{}_{12}$ is very sensitive to small perturbations to model parameters in the region around $\tau = {\rm i}$, radiative corrections via RG running to the model parameters should be important. Given the same input parameters and ${\rm Re}\,\tau \sim -0.03$, the mixing angle $\theta^{}_{12}$ would be changed from $32^\circ$ to $50^\circ_{}$ after the radiative corrections are taken into consideration. The main reason for such a significant RG running effect is that the mass degeneracy between $m^{}_1$ and $m^{}_2$ is governed by small perturbation parameters while $\theta^{}_{12}$ is determined by the ratio of two small perturbation parameters. Although the details of such an analysis in this paper are quite model-dependent, it does shed some light on the common features of the modular-symmetry models that seem to be ``unstable'' around the stabilizers. For instance, neutrino masses are usually predicted to be nearly degenerate in such kinds of models.

In the near future, it is necessary and interesting to further investigate the properties of modular-symmetry models around all the stabilizers in a systematic and model-independent way. As such scenarios are strictly constrained by the modular symmetry and the residual symmetries, their predictions for neutrino masses, mixing angles and the Dirac CP-violating phase are readily to be tested in next-generation neutrino oscillation experiments, neutrinoless double-beta decays and cosmological observations.

\section*{Acknowledgements}
This work was supported in part by the National Natural Science Foundation of China under grant No.~11775232 and No.~11835013, by the Key Research Program of the Chinese Academy of
Sciences under grant No. XDPB15, and by the CAS Center for Excellence in Particle Physics.

\newpage

\appendix

\section{Modular $A^\prime_5$ group and the modular space ${\cal M}^{}_1 \left[\Gamma(5)\right]$}\label{app:A}

The $A^\prime_5$ group has 120 elements, which can be divided into nine conjugacy classes, indicating that $A^{\prime}_{5}$ has nine distinct irreducible representations that are normally denoted as ${\bf 1}$, $\widehat{\bf 2}$, $\widehat{\bf 2}^{\prime}_{}$, ${\bf 3}$, ${\bf 3}^{\prime}_{}$, ${\bf 4}$, $\widehat{\bf 4}$, ${\bf 5}$ and $\widehat{\bf 6}$ by their dimensions. The representation matrices of all three generators $S$, $T$ and $R$ in the irreducible representations are summarized as below
\begin{eqnarray}
{\bf 1} &: & \rho(S) = + 1 \; , \quad \rho(T) = + 1 \; , \quad \rho(R) = +1 \; , \nonumber \\
\widehat{\bf 2} &: & \rho(S)=\dfrac{\rm i}{\sqrt[4]{5}}\left(
\begin{matrix}
\sqrt{\phi} & \sqrt{\phi-1} \\
\sqrt{\phi-1} & -\sqrt{\phi} \\
\end{matrix}\right) \; , \quad
\rho(T)=\left(
\begin{matrix}
\omega^2_{} & 0 \\
0 & \omega^3_{} \\
\end{matrix}\right) \; , \quad
\rho(R)=- \mathbb{I}^{}_{2\times 2}\; , \nonumber \\
\widehat{\bf 2}^{\prime}_{} &: & \rho(S)=\dfrac{\rm i}{\sqrt[4]{5}}\left(
\begin{matrix}
\sqrt{\phi-1} & \sqrt{\phi} \\
\sqrt{\phi} & -\sqrt{\phi-1} \\
\end{matrix}\right) \; , \quad
\rho(T)=\left(
\begin{matrix}
\omega & 0 \\
0 & \omega^4_{} \\
\end{matrix}\right) \; , \quad
\rho(R)=- \mathbb{I}^{}_{2\times 2}\; , \nonumber  \\
{\bf 3} &:& \rho(S)=\dfrac{1}{\sqrt{5}}\left(\begin{matrix}
1 & -\sqrt{2} & -\sqrt{2} \\
-\sqrt{2} & -\phi & \phi-1 \\
-\sqrt{2} & \phi-1 & -\phi \\
\end{matrix}\right) \; , \quad \rho(T)=\left(\begin{matrix}
1 & 0 & 0 \\ 0 & \omega & 0 \\ 0 & 0 & \omega^4_{}
\end{matrix}\right) \; ,  \quad \rho(R)=+\mathbb{I}^{}_{3\times 3}\; , \nonumber  \\
{\bf 3}^{\prime}_{} &: & \rho(S)=\dfrac{1}{\sqrt{5}}\left(\begin{matrix}
-1 & \sqrt{2} & \sqrt{2} \\
\sqrt{2} & 1-\phi & \phi \\
\sqrt{2} & \phi & 1-\phi \\
\end{matrix}\right) \; , \quad \rho(T)=\left(\begin{matrix}
1 & 0 & 0 \\
0 & \omega^2_{} & 0 \\
0 & 0 & \omega^3_{} \\
\end{matrix}\right) \; , \quad \rho(R)=+\mathbb{I}^{}_{3\times 3}\; , \nonumber  \\
{\bf 4} &:& \rho(S) =  \dfrac{1}{\sqrt{5}}
\left(\begin{matrix}
1 & \phi-1 & \phi & -1 \\
\phi-1 & -1 & 1 & \phi \\
\phi & 1 & -1 & \phi-1 \\
-1 & \phi & \phi-1 & 1 \\
\end{matrix}\right) \; , \quad
\rho(T)=\left(\begin{matrix}
\omega & 0 & 0 & 0\\
0 & \omega^2_{} & 0 &  0\\
0 & 0 & \omega^3_{} &  0 \\
0 & 0 & 0 & \omega^4_{} \\
\end{matrix}\right)  \; , \quad \rho(R)=+\mathbb{I}^{}_{4\times 4}\; , \nonumber  \\
\widehat{\bf 4} &:& \rho(S) =  \dfrac{\rm i}{5^{\frac{3}{4}}_{}}
\left(\begin{matrix}
-\sqrt{2\phi+1} & \sqrt{3\phi} & \sqrt{3(\phi-1)} & \sqrt{2\phi-3} \\
\sqrt{3\phi} & \sqrt{2\phi-3} & \sqrt{2\phi+1} & \sqrt{3(\phi-1)} \\
\sqrt{3(\phi-1)} & \sqrt{2\phi+1} & -\sqrt{2\phi-3} & -\sqrt{3\phi} \\
\sqrt{2\phi-3} & \sqrt{3(\phi-1)} & -\sqrt{3\phi} & \sqrt{2\phi+1} \\
\end{matrix}\right) \; , \nonumber \\
&&\rho(T)=\left(\begin{matrix}
\omega & 0 & 0 & 0\\
0 & \omega^2_{} & 0 &  0\\
0 & 0 & \omega^3_{} &  0 \\
0 & 0 & 0 & \omega^4_{} \\
\end{matrix}\right)  \; , \quad
\rho(R)=-\mathbb{I}^{}_{4\times 4}\; , \nonumber \\
{\bf 5} &:& \rho(S) =  \dfrac{1}{5}
\left(\begin{matrix}
-1 & \sqrt{6} & \sqrt{6} & \sqrt{6} & \sqrt{6} \\
\sqrt{6} & (\phi-1)^2_{} & -2\phi & 2(\phi-1) & \phi^2_{}\\
\sqrt{6} & -2\phi & \phi^2_{} & (\phi-1)^2_{} & 2(\phi-1) \\
\sqrt{6} & 2(\phi-1) & (\phi-1)^2_{} & \phi^2_{} & -2\phi \\
\sqrt{6} & \phi^2_{} & 2(\phi-1) & -2\phi & (\phi-1)^2_{}
\end{matrix}\right) \; , \nonumber \\
&&\rho(T)=\left(\begin{matrix}
1 & 0 & 0 & 0 & 0 \\
0 & \omega & 0 & 0 & 0\\
0 & 0 & \omega^2_{} & 0 &  0\\
0 & 0 & 0 & \omega^3_{} &  0 \\
0 & 0 & 0 & 0 & \omega^4_{} \\
\end{matrix}\right)  \; , \quad
 \rho(R)=+\mathbb{I}^{}_{5\times 5}\; ,  \nonumber
\label{eq:irrep1}
\end{eqnarray}
\begin{eqnarray}
\widehat{\bf 6} &:&   \rho(S)=\dfrac{-\rm i}{5^{\frac{3}{4}}_{}}
\left(\begin{matrix}
\sqrt{\phi} &-\sqrt{2(\phi-1)} &\sqrt{2\phi-3} &-\sqrt{2\phi+1} & \sqrt{2\phi}& \sqrt{\phi-1} \\
-\sqrt{2(\phi-1)} &-\sqrt{\phi-1} &\sqrt{2(\phi-1)} &\sqrt{2\phi} &\sqrt{\phi}  &\sqrt{2\phi}\\
\sqrt{2\phi-3} & \sqrt{2(\phi-1)} & \sqrt{\phi} & -\sqrt{\phi-1}& -\sqrt{2\phi} & \sqrt{2\phi+1} \\
-\sqrt{2\phi+1} & \sqrt{2\phi} & -\sqrt{\phi-1} & -\sqrt{\phi} & \sqrt{2(\phi-1)} & \sqrt{2\phi-3} \\
\sqrt{2\phi} & \sqrt{\phi} & -\sqrt{2\phi}& \sqrt{2(\phi-1)} & \sqrt{\phi-1} &  \sqrt{2(\phi-1)} \\
\sqrt{\phi-1} & \sqrt{2\phi} & \sqrt{2\phi+1} &  \sqrt{2\phi-3} &  \sqrt{2(\phi-1)} & -\sqrt{\phi} \\
\end{matrix} \right) \; ,  \nonumber   \\
& & \rho(T) =\left(\begin{matrix}
1 & 0 & 0 & 0 & 0 & 0\\
0 &  \omega & 0 & 0 & 0 & 0\\
0 &  0 & \omega^2_{} & 0  & 0 & 0\\
0 & 0 & 0 & \omega^3_{} & 0  & 0 \\
0 & 0 & 0 & 0 & \omega^4_{} & 0  \\
0 & 0 & 0 & 0 & 0 & 1\\
\end{matrix}\right) \; ,  \quad \rho(R) = -\mathbb{I}^{}_{6\times 6} \; , \nonumber
\label{eq:irrep2}
\end{eqnarray}
where $\omega \equiv e^{2{\rm i}\pi/5}$ and $\mathbb{I}^{}_{n\times n}$ denotes the $n$-dimensional identity matrix. Notice that the representations ${\bf 1}$, ${\bf 3}$, ${\bf 3}^{\prime}_{}$, ${\bf 4}$ and ${\bf 5}$ with $R = \mathbb{I}$ coincide with those for $A^{}_{5}$, whereas $\widehat{\bf 2}$, $\widehat{\bf 2}^{\prime}_{}$, $\widehat{\bf 4}$ and $\widehat{\bf 6}$ are unique for $A^\prime_5$ with $R = -\mathbb{I}$.

The whole list of decomposition rules for the Kronecker products of any two nontrivial irreducible representations of $A^{\prime}_5$ can be found in Ref.~\cite{Wang:2020lxk}. Here we only summarize the decomposition rules relevant to this paper.
\begin{itemize}
	\item $\widehat{\bf 2} \otimes \widehat{\bf 2} = {\bf 1}^{}_{\rm a} \oplus {\bf 3}^{}_{\rm s}$
	\begin{eqnarray}
	{\bf 1}^{}_{\rm a} : - \dfrac{\sqrt{2}}{2} \left(\alpha_1^{ } \beta_2^{ } - \alpha_2^{ } \beta_1^{ }\right) \; ,\quad
	{\bf 3}^{}_{\rm s} : \dfrac{\sqrt{2}}{2} \left(
	\begin{array}{c}
	\alpha _1^{ } \beta
	_2^{ }+\alpha _2^{ } \beta _1^{ } \\
	-\sqrt{2} \alpha _2^{ } \beta _2^{ } \\
	\sqrt{2} \alpha _1^{ } \beta _1^{ } \\
	\end{array}
	\right) \; . \nonumber
	\end{eqnarray}
	\item $\widehat{\bf 2} \otimes {\bf 3}^{\prime}_{} = \widehat{\bf 6}$
	\begin{eqnarray}
	\begin{array}{c}
	\widehat{\bf 6} : -\dfrac{\sqrt{2}}{2}\left(
	\begin{array}{c}
	\alpha _1^{ } \beta
	_3^{ }-\alpha _2^{ } \beta _2^{ } \\
	\sqrt{2} \alpha _2^{ } \beta _3^{ } \\
	-\sqrt{2} \alpha _1^{ } \beta _1^{ } \\
	\sqrt{2} \alpha _2^{ } \beta _1^{ } \\
	-\sqrt{2} \alpha _1^{ } \beta _2^{ } \\
	\alpha _1^{ } \beta
	_3^{ }+\alpha _2^{ } \beta _2^{ } \\
	\end{array}
	\right) \; . \nonumber
	\end{array}
	\end{eqnarray}
	\item $\mathbf{3} \otimes \boldsymbol{3}=\mathbf{1}^{}_{\rm s} \oplus \boldsymbol{3}^{}_{\rm a} \oplus \boldsymbol{5}^{}_{\rm s}$
	\begin{eqnarray}
	\begin{array}{c}
	\mathbf{1}^{}_{\rm s} : \dfrac{\sqrt{3}}{3} (\alpha^{}_{1} \beta^{}_{1}+\alpha^{}_{2} \beta^{}_{3}+\alpha^{}_{3} \beta^{}_{2}) \; , \\
	{\bf 3}^{}_{\rm a} : \dfrac{\sqrt{2}}{2} \left(\begin{array}{c}
	\alpha^{}_{2} \beta^{}_{3}-\alpha^{}_{3} \beta^{}_{2} \\
	\alpha^{}_{1} \beta^{}_{2}-\alpha^{}_{2} \beta^{}_{1} \\
	\alpha^{}_{3} \beta^{}_{1}-\alpha^{}_{1} \beta^{}_{3}
	\end{array}\right) \; , \quad
	{\bf 5}^{}_{\rm s} : \dfrac{\sqrt{6}}{6} \left(\begin{array}{c}
	2 \alpha^{}_{1} \beta^{}_{1}-\alpha^{}_{2} \beta^{}_{3}-\alpha^{}_{3} \beta^{}_{2} \\
	-\sqrt{3} \alpha^{}_{1} \beta^{}_{2}-\sqrt{3} \alpha^{}_{2} \beta^{}_{1} \\
	\sqrt{6} \alpha^{}_{2} \beta^{}_{2} \\
	\sqrt{6} \alpha^{}_{3} \beta^{}_{3} \\
	-\sqrt{3} \left(\alpha^{}_{1} \beta^{}_{3}+ \alpha^{}_{3} \beta^{}_{1} \right)
	\end{array}\right) \; .
	\end{array} \nonumber
	\end{eqnarray}
	\item ${\bf 3}^{\prime}_{} \otimes {\bf 3}^{\prime}_{} = {\bf 1}^{}_{\rm s} \oplus {\bf 3}^{\prime}_{\rm a} \oplus {\bf 5}^{}_{\rm s}$
	\begin{eqnarray}
	\begin{array}{c}
	{\bf 1}^{}_{\rm s} : \dfrac{\sqrt{3}}{3} \left(\alpha^{}_1 \beta^{}_1 + \alpha^{}_2 \beta^{}_3 + \alpha^{}_3 \beta^{}_2\right) \; ,\\
	{\bf 3}^{\prime}_{\rm a} : \dfrac{\sqrt{2}}{2} \left(\begin{array}{c}
	\alpha^{}_{2} \beta^{}_{3}-\alpha^{}_{3} \beta^{}_{2} \\
	\alpha^{}_{1} \beta^{}_{2}-\alpha^{}_{2} \beta^{}_{1} \\
	\alpha^{}_{3} \beta^{}_{1}-\alpha^{}_{1} \beta^{}_{3}
	\end{array}\right) \; ,\quad
	{\bf 5}^{}_{\rm s} : \dfrac{\sqrt{6}}{6} \left(\begin{array}{c}
	2 \alpha^{}_{1} \beta^{}_{1}-\alpha^{}_{2} \beta^{}_{3}-\alpha^{}_{3} \beta^{}_{2} \\
	\sqrt{6} \alpha^{}_{3} \beta^{}_{3} \\
	-\sqrt{3} \left(\alpha^{}_{1} \beta^{}_{2}+ \alpha^{}_{2} \beta^{}_{1} \right) \\
	-\sqrt{3} \left( \alpha^{}_{1} \beta^{}_{3}+ \alpha^{}_{3} \beta^{}_{1} \right) \\
	\sqrt{6} \alpha^{}_{2} \beta^{}_{2}
	\end{array}\right) \; .
	\end{array} \nonumber
	\end{eqnarray}
		\item ${\bf 5} \otimes {\bf 5} = {\bf 1}^{}_{\rm s} \oplus {\bf 3}^{}_{\rm a} \oplus {\bf 3}^{\prime}_{\rm a} \oplus {\bf 4}^{}_{\rm s} \oplus {\bf 4}^{}_{\rm a} \oplus {\bf 5}^{}_{\rm s,1} \oplus {\bf 5}^{}_{\rm s,2} $
	\begin{eqnarray}
	\begin{array}{c}
	{\bf 1}^{}_{\rm s} : \dfrac{\sqrt{5}}{5} \left(\alpha^{}_{1} \beta^{}_{1}+\alpha^{}_{2} \beta^{}_{5}+\alpha^{}_{3} \beta^{}_{4}+\alpha^{}_{4} \beta^{}_{3}+\alpha^{}_{5} \beta^{}_{2}\right) \; , \\
	{\bf 3}^{}_{\rm a} : \dfrac{\sqrt{10}}{10} \left(\begin{array}{c}
	\alpha^{}_{2} \beta^{}_{5}+2 \alpha^{}_{3} \beta^{}_{4}-2 \alpha^{}_{4} \beta^{}_{3}-\alpha^{}_{5} \beta^{}_{2}  \\
	-\sqrt{3} \alpha^{}_{1} \beta^{}_{2}+\sqrt{3} \alpha^{}_{2} \beta^{}_{1}+\sqrt{2} \alpha^{}_{3} \beta^{}_{5}-\sqrt{2} \alpha^{}_{5} \beta^{}_{3} \\
	\sqrt{3} \alpha^{}_{1} \beta^{}_{5}+\sqrt{2} \alpha^{}_{2} \beta^{}_{4}-\sqrt{2} \alpha^{}_{4} \beta^{}_{2}-\sqrt{3} \alpha^{}_{5} \beta^{}_{1}
	\end{array}\right) \; , \\
	{\bf 3}^{\prime}_{\rm a} : \dfrac{\sqrt{10}}{10} \left(\begin{array}{c}
	2 \alpha^{}_{2} \beta^{}_{5}-\alpha^{}_{3} \beta^{}_{4}+\alpha^{}_{4} \beta^{}_{3}-2 \alpha^{}_{5} \beta^{}_{2} \\
	\sqrt{3} \alpha^{}_{1} \beta^{}_{3}-\sqrt{3} \alpha^{}_{3} \beta^{}_{1}+\sqrt{2} \alpha^{}_{4} \beta^{}_{5}-\sqrt{2} \alpha^{}_{5} \beta^{}_{4} \\
	-\sqrt{3} \alpha^{}_{1} \beta^{}_{4}+\sqrt{2} \alpha^{}_{2} \beta^{}_{3}-\sqrt{2} \alpha^{}_{3} \beta^{}_{2}+\sqrt{3} \alpha^{}_{4} \beta^{}_{1}
	\end{array}\right) \; , \\
	{\bf 4}^{}_{\rm s} : \dfrac{\sqrt{30}}{30} \left(\begin{array}{l}
	\sqrt{6} \alpha^{}_{1} \beta^{}_{2}+\sqrt{6} \alpha^{}_{2} \beta^{}_{1}- \alpha^{}_{3} \beta^{}_{5}+4  \alpha^{}_{4} \beta^{}_{4}- \alpha^{}_{5} \beta^{}_{3} \\
	\sqrt{6} \alpha^{}_{1} \beta^{}_{3}+4  \alpha^{}_{2} \beta^{}_{2}+\sqrt{6} \alpha^{}_{3} \beta^{}_{1}- \alpha^{}_{4} \beta^{}_{5}- \alpha^{}_{5} \beta^{}_{4} \\
	\sqrt{6} \alpha^{}_{1} \beta^{}_{4}- \alpha^{}_{2} \beta^{}_{3}- \alpha^{}_{3} \beta^{}_{2}+\sqrt{6} \alpha^{}_{4} \beta^{}_{1}+4  \alpha^{}_{5} \beta^{}_{5} \\
	\sqrt{6} \alpha^{}_{1} \beta^{}_{5}- \alpha^{}_{2} \beta^{}_{4}+4  \alpha^{}_{3} \beta^{}_{3}- \alpha^{}_{4} \beta^{}_{2}+\sqrt{6} \alpha^{}_{5} \beta^{}_{1}
	\end{array}\right)\; ,\\
	{\bf 4}^{}_{\rm a} : \dfrac{\sqrt{10}}{10} \left(\begin{array}{c}
	\sqrt{2} \alpha^{}_{1} \beta^{}_{2}-\sqrt{2} \alpha^{}_{2} \beta^{}_{1}+\sqrt{3} \alpha^{}_{3} \beta^{}_{5}-\sqrt{3} \alpha^{}_{5} \beta^{}_{3} \\
	-\sqrt{2} \alpha^{}_{1} \beta^{}_{3}+\sqrt{2} \alpha^{}_{3} \beta^{}_{1}+\sqrt{3} \alpha^{}_{4} \beta^{}_{5}-\sqrt{3} \alpha^{}_{5} \beta^{}_{4} \\
	-\sqrt{2} \alpha^{}_{1} \beta^{}_{4}-\sqrt{3} \alpha^{}_{2} \beta^{}_{3}+\sqrt{3} \alpha^{}_{3} \beta^{}_{2}+\sqrt{2} \alpha^{}_{4} \beta^{}_{1} \\
	\sqrt{2} \alpha^{}_{1} \beta^{}_{5}-\sqrt{3} \alpha^{}_{2} \beta^{}_{4}+\sqrt{3} \alpha^{}_{4} \beta^{}_{2}-\sqrt{2} \alpha^{}_{5} \beta^{}_{1}
	\end{array}\right) \; , \\
	{\bf 5}^{}_{\rm s,1} : \dfrac{\sqrt{14}}{14} \left(\begin{array}{c}
	2 \alpha^{}_{1} \beta^{}_{1}+\alpha^{}_{2} \beta^{}_{5}-2 \alpha^{}_{3} \beta^{}_{4}-2 \alpha^{}_{4} \beta^{}_{3}+\alpha^{}_{5} \beta^{}_{2} \\
	\alpha^{}_{1} \beta^{}_{2}+\alpha^{}_{2} \beta^{}_{1}+\sqrt{6} \alpha^{}_{3} \beta^{}_{5}+\sqrt{6} \alpha^{}_{5} \beta^{}_{3} \\
	-2 \alpha^{}_{1} \beta^{}_{3}+\sqrt{6} \alpha^{}_{2} \beta^{}_{2}-2 \alpha^{}_{3} \beta^{}_{1} \\
	-2 \alpha^{}_{1} \beta^{}_{4}-2 \alpha^{}_{4} \beta^{}_{1}+\sqrt{6} \alpha^{}_{5} \beta^{}_{5} \\
	\alpha^{}_{1} \beta^{}_{5}+\sqrt{6} \alpha^{}_{2} \beta^{}_{4}+\sqrt{6} \alpha^{}_{4} \beta^{}_{2}+\alpha^{}_{5} \beta^{}_{1}
	\end{array}\right) \; ,  \\
	{\bf 5}^{}_{\rm s,2} : \dfrac{\sqrt{14}}{14} \left(\begin{array}{c}
	2 \alpha_{1} \beta_{1}-2 \alpha_{2} \beta_{5}+\alpha_{3} \beta_{4}+\alpha_{4} \beta_{3}-2 \alpha_{5} \beta_{2} \\
	-2 \alpha_{1} \beta_{2}-2 \alpha_{2} \beta_{1}+\sqrt{6} \alpha_{4} \beta_{4} \\
	\alpha_{1} \beta_{3}+\alpha_{3} \beta_{1}+\sqrt{6} \alpha_{4} \beta_{5}+\sqrt{6} \alpha_{5} \beta_{4} \\
	\alpha_{1} \beta_{4}+\sqrt{6} \alpha_{2} \beta_{3}+\sqrt{6} \alpha_{3} \beta_{2}+\alpha_{4} \beta_{1} \\
	-2 \alpha_{1} \beta_{5}+\sqrt{6} \alpha_{3} \beta_{3}-2 \alpha_{5} \beta_{1}
	\end{array}\right) \; .
	\end{array} \nonumber
	\end{eqnarray}
	\item $\widehat{\bf 6} \otimes \widehat{\bf 6} = {\bf 1}^{}_{\rm a} \oplus {\bf 3}^{}_{\rm s,1} \oplus {\bf 3}^{}_{\rm s,2} \oplus {\bf 3}^{\prime}_{\rm s,1} \oplus {\bf 3}^{\prime}_{\rm s,2} \oplus {\bf 4}^{}_{\rm s} \oplus {\bf 4}^{}_{\rm a} \oplus {\bf 5}^{}_{\rm s} \oplus {\bf 5}^{}_{\rm a,1} \oplus {\bf 5}^{}_{\rm a,2}$
	\begin{eqnarray}
	\begin{array}{c}
	{\bf 1}^{}_{\rm a} : \dfrac{\sqrt{6}}{6} \left(\alpha_{1}^{} \beta_{6}^{}+\alpha_{2}^{} \beta_{5}^{}-\alpha_{3}^{} \beta_{4}^{}+\alpha_{4}^{} \beta_{3}^{}-\alpha_{5}^{} \beta_{2}^{}-\alpha_{6}^{} \beta_{1}^{}\right) \; ,\\
	{\bf 3}^{}_{\rm s,1} : \dfrac{1}{2}\left(\begin{array}{l}
	\alpha_{1}^{} \beta_{6}^{}+\alpha_{3}^{} \beta_{4}^{}+\alpha_{4}^{} \beta_{3}^{}+\alpha_{6}^{} \beta_{1}^{} \\
	\alpha_{2}^{} \beta_{6}^{}+\alpha_{3}^{} \beta_{5}^{}+\alpha_{5}^{} \beta_{3}^{}+\alpha_{6}^{} \beta_{2}^{} \\
	\alpha_{1}^{} \beta_{5}^{}-\alpha_{2}^{} \beta_{4}^{}-\alpha_{4}^{} \beta_{2}^{}+\alpha_{5}^{} \beta_{1}^{} 	
	\end{array}\right) \; , \nonumber
	\end{array}
	\end{eqnarray}
	\begin{eqnarray}
	\begin{array}{c}
	{\bf 3}^{}_{\rm s,2} : \dfrac{\sqrt{6}}{6}\left(\begin{array}{c}
	-\alpha_{1}^{} \beta_{1}^{}+\alpha_{1}^{} \beta_{6}^{}-\alpha_{2}^{} \beta_{5}^{}-\alpha_{5}^{} \beta_{2}^{}+\alpha_{6}^{} \beta_{1}^{}+\alpha_{6}^{} \beta_{6}^{} \\
	\alpha_{1}^{} \beta_{2}^{}+\alpha_{2}^{} \beta_{1}^{}+\alpha_{3}^{} \beta_{5}^{}-\sqrt{2} \alpha_{4}^{} \beta_{4}^{}+\alpha_{5}^{} \beta_{3}^{} \\
	-\alpha_{2}^{} \beta_{4}^{}+\sqrt{2} \alpha_{3}^{} \beta_{3}^{}-\alpha_{4}^{} \beta_{2}^{}-\alpha_{5}^{} \beta_{6}^{}-\alpha_{6}^{} \beta_{5}^{}
	\end{array}\right) \; , \\
	{\bf 3}^{\prime}_{\rm s,1} : \dfrac{\sqrt{6}}{6}\left(\begin{array}{c}
	\alpha_{1}^{} \beta_{6}^{}+\alpha_{2}^{} \beta_{5}^{}-\alpha_{3}^{} \beta_{4}^{}-\alpha_{4}^{} \beta_{3}^{}+\alpha_{5}^{} \beta_{2}^{}+\alpha_{6}^{} \beta_{1}^{} \\
	-\sqrt{2}\left(\alpha_{1}^{} \beta_{3}^{}-\alpha_{2}^{} \beta_{2}^{}+\alpha_{3}^{} \beta_{1}^{}\right) \\
	-\sqrt{2}\left(\alpha_{4}^{} \beta_{6}^{}+\alpha_{5}^{} \beta_{5}^{}+\alpha_{6}^{} \beta_{4}^{}\right)
	\end{array}\right) \; , \\
	{\bf 3}^{\prime}_{\rm s,2} : \dfrac{\sqrt{2}}{4}\left(\begin{array}{c}
	\sqrt{2}\left(\alpha_{1}^{} \beta_{1}^{}-\alpha_{2}^{} \beta_{5}^{}-\alpha_{5}^{} \beta_{2}^{}-\alpha_{6}^{} \beta_{6}^{}\right) \\
	\alpha_{1}^{} \beta_{3}^{}+\alpha_{3}^{} \beta_{1}^{}-\alpha_{3}^{} \beta_{6}^{}+\sqrt{2} \alpha_{4}^{} \beta_{5}^{}+\sqrt{2} \alpha_{5}^{} \beta_{4}^{}-\alpha_{6}^{} \beta_{3}^{} \\
	\alpha_{1}^{} \beta_{4}^{}+\sqrt{2} \alpha_{2}^{} \beta_{3}^{}+\sqrt{2} \alpha_{3}^{} \beta_{2}^{}+\alpha_{4}^{} \beta_{1}^{}+\alpha_{4}^{} \beta_{6}^{}+\alpha_{6}^{} \beta_{4}^{}
	\end{array}\right) \; , \\
	{\bf 4}^{}_{\rm s} : \dfrac{\sqrt{30}}{30}\left(\begin{array}{c}
	\begin{array}{c}
	-3 \alpha_{1} \beta_{2}-3  \alpha_{2} \beta_{1}- \alpha_{2} \beta_{6}+ \alpha_{3} \beta_{5}-2 \sqrt{2} \alpha_{4} \beta_{4}+\alpha_{5} \beta_{3}-\alpha_{6} \beta_{2} \\
	\sqrt{2}\alpha_{1} \beta_{3}+2\sqrt{2} \alpha_{2} \beta_{2}+\sqrt{2} \alpha_{3} \beta_{1}+2\sqrt{2} \alpha_{3} \beta_{6}+ \alpha_{4} \beta_{5}+ \alpha_{5} \beta_{4}+2\sqrt{2} \alpha_{6} \beta_{3} \\
	2\sqrt{2} \alpha_{1} \beta_{4}- \alpha_{2} \beta_{3}- \alpha_{3} \beta_{2}+2\sqrt{2} \alpha_{4} \beta_{1}-\sqrt{2} \alpha_{4} \beta_{6}+2\sqrt{2} \alpha_{5} \beta_{5}-\sqrt{2} \alpha_{6} \beta_{4} \\
	\alpha_{1} \beta_{5}+ \alpha_{2} \beta_{4}-2\sqrt{2}\alpha_{3} \beta_{3}+ \alpha_{4} \beta_{2}+\alpha_{5} \beta_{1}-3 \alpha_{5} \beta_{6}-3  \alpha_{6} \beta_{5}
	\end{array}
	\end{array}\right) \; , \\
	{\bf 4}^{}_{\rm a} :-\dfrac{\sqrt{6}}{6}\left(\begin{array}{c}
	\begin{array}{c}
	\alpha_{1}^{} \beta_{2}^{}-\alpha_{2}^{} \beta_{1}^{}+\alpha_{2}^{} \beta_{6}^{}+\alpha_{3}^{} \beta_{5}^{}-\alpha_{5}^{} \beta_{3}^{}-\alpha_{6}^{} \beta_{2}^{} \\
	\sqrt{2} \alpha_{1}^{} \beta_{3}^{}-\sqrt{2} \alpha_{3}^{} \beta_{1}^{}+\alpha_{4}^{} \beta_{5}^{}-\alpha_{5}^{} \beta_{4}^{} \\
	\alpha_{2}^{} \beta_{3}^{}-\alpha_{3}^{} \beta_{2}^{}+\sqrt{2}\alpha_{4}^{} \beta_{6}^{}-\sqrt{2}\alpha_{6}^{} \beta_{4}^{} \\
	\alpha_{1}^{} \beta_{5}^{}-\alpha_{2}^{} \beta_{4}^{}+\alpha_{4}^{} \beta_{2}^{}-\alpha_{5}^{} \beta_{1}^{}-\alpha_{5}^{} \beta_{6}^{}+\alpha_{6}^{} \beta_{5}^{}
	\end{array}
	\end{array}\right) \; , \\
	{\bf 5}^{}_{\rm s} : \dfrac{ \sqrt{3}}{6} \left( \begin{array}{c}
	\sqrt{6}\left(\alpha^{}_{1} \beta^{}_{1}+\alpha^{}_{6} \beta^{}_{6}\right) \\
	\sqrt{2} \left( \alpha^{}_{2} \beta^{}_{6}- \alpha^{}_{3} \beta^{}_{5}-\sqrt{2} \alpha^{}_{4} \beta^{}_{4}- \alpha^{}_{5} \beta^{}_{3}+ \alpha^{}_{6} \beta^{}_{2} \right) \\
	-\alpha^{}_{1} \beta^{}_{3}-2 \alpha^{}_{2} \beta^{}_{2}-\alpha^{}_{3} \beta^{}_{1}+\alpha^{}_{3} \beta^{}_{6}+\sqrt{2} \alpha^{}_{4} \beta^{}_{5}+\sqrt{2} \alpha^{}_{5} \beta^{}_{4}+\alpha^{}_{6} \beta^{}_{3} \\
	\alpha^{}_{1} \beta^{}_{4}-\sqrt{2} \alpha^{}_{2} \beta^{}_{3}-\sqrt{2} \alpha^{}_{3} \beta^{}_{2}+\alpha^{}_{4} \beta^{}_{1}+\alpha^{}_{4} \beta^{}_{6}-2 \alpha^{}_{5} \beta^{}_{5}+\alpha^{}_{6} \beta^{}_{4} \\
	-\sqrt{2} \left( \alpha^{}_{1} \beta^{}_{5}+ \alpha^{}_{2} \beta^{}_{4}+ \sqrt{2}  \alpha^{}_{3} \beta^{}_{3}+ \alpha^{}_{4} \beta^{}_{2} + \alpha^{}_{5} \beta^{}_{1} \right)
	\end{array}\right) \; .\\
	{\bf 5}^{}_{\rm a,1} : \dfrac{\sqrt{3}}{6}\left(\begin{array}{c}
	\alpha^{}_{1} \beta^{}_{6}-2 \alpha^{}_{2} \beta^{}_{5}-\alpha^{}_{3} \beta^{}_{4}+\alpha^{}_{4} \beta^{}_{3}+2 \alpha^{}_{5} \beta^{}_{2}-\alpha^{}_{6} \beta^{}_{1} \\
	-\sqrt{3}\left(\alpha^{}_{2} \beta^{}_{6}-\alpha^{}_{3} \beta^{}_{5}+\alpha^{}_{5} \beta^{}_{3}-\alpha^{}_{6} \beta^{}_{2}\right) \\
	\sqrt{6}\left(\alpha^{}_{3} \beta^{}_{6}-\alpha^{}_{6} \beta^{}_{3}\right) \\
	-\sqrt{6}\left(\alpha^{}_{1} \beta^{}_{4}-\alpha^{}_{4} \beta^{}_{1}\right) \\
	-\sqrt{3}\left(\alpha^{}_{1} \beta^{}_{5}+\alpha^{}_{2} \beta^{}_{4}-\alpha^{}_{4} \beta^{}_{2}-\alpha^{}_{5} \beta^{}_{1}\right)
	\end{array}\right)  \; , \\
	{\bf 5}^{}_{\rm a,2} : \dfrac{\sqrt{3}}{6} \left( \begin{array}{c}
	\sqrt{3}\left(\alpha^{}_{1} \beta^{}_{6}+\alpha^{}_{3} \beta^{}_{4}-\alpha^{}_{4} \beta^{}_{3}-\alpha^{}_{6} \beta^{}_{1}\right) \\
	-2 \alpha^{}_{1} \beta^{}_{2}+2 \alpha^{}_{2} \beta^{}_{1}+\alpha^{}_{2} \beta^{}_{6}+\alpha^{}_{3} \beta^{}_{5}-\alpha^{}_{5} \beta^{}_{3}-\alpha^{}_{6} \beta^{}_{2} \\
	-\sqrt{2} \left( \alpha^{}_{1} \beta^{}_{3}- \alpha^{}_{3} \beta^{}_{1}-\sqrt{2} \alpha^{}_{4} \beta^{}_{5}+ \sqrt{2} \alpha^{}_{5} \beta^{}_{4}  \right) \\
	\sqrt{2} \left( \sqrt{2} \alpha^{}_{2} \beta^{}_{3}-\sqrt{2} \alpha^{}_{3} \beta^{}_{2}-\alpha^{}_{4} \beta^{}_{6}+\alpha^{}_{6} \beta^{}_{4}\right) \\
	\alpha^{}_{1} \beta^{}_{5}-\alpha^{}_{2} \beta^{}_{4}+\alpha^{}_{4} \beta^{}_{2}-\alpha^{}_{5} \beta^{}_{1}+2 \alpha^{}_{5} \beta^{}_{6}-2 \alpha^{}_{6} \beta^{}_{5}
	\end{array} \right) \; .
	\end{array} \nonumber
	\end{eqnarray}
\end{itemize}

For a given non-negative integer $k$, the modular space ${\cal M}^{}_{k}\left[\Gamma(5)\right]$ of weight $k$ for $\Gamma(5)$ contains $5k+1$ linearly-independent modular forms, which can be regarded as the basis vectors of the modular space. According to Ref.~\cite{Schultz:2015}, we have
\begin{eqnarray}
{\cal M}^{}_k\left[\Gamma(5)\right] = \bigoplus^{}_{\substack{a+b = 5k \\a, b \geq 0}} \mathbb{C}\, \frac{\eta(5 \tau)^{15 k}}{\eta(\tau)^{3 k}} \, {\mathfrak k}^a_{\frac{1}{5},\frac{0}{5}}(5\tau) \, {\mathfrak k}^b_{\frac{2}{5},\frac{0}{5}}(5\tau) \; ,
\label{eq:G5basis}
%     (A.1)
\end{eqnarray}
where $\eta(\tau)$ is the Dedekind eta function
\begin{eqnarray}
\eta(\tau)=q^{1 / 24}_{} \prod_{n=1}^{\infty}\left(1-q^{n}_{}\right) \; ,
\label{eq:Dedekindeta}
%     (A.2)
\end{eqnarray}
with $q \equiv e^{2 {\rm i} \pi \tau}_{}$, and ${\mathfrak k}^{}_{r^{}_1,r^{}_2}(\tau)$ is the Klein form
\begin{eqnarray}
\mathfrak{k}^{}_{r^{}_{1}, r^{}_{2}}(\tau)= q_{z}^{(r^{}_{1}-1) / 2}\left(1-q^{}_{z}\right) \times \prod_{n=1}^{\infty}\left(1-q^{n}_{} q^{}_{z}\right)\left(1-q^{n}_{} q_{z}^{-1}\right)\left(1-q^{n}_{}\right)^{-2}_{} \; ,
\label{eq:Kexpansion}
%     (A.3)
\end{eqnarray}
with $(r^{}_1,r^{}_2)$ being a pair of rational numbers in the domain of ${\mathbb Q}^2_{}-{\mathbb Z}^2_{}$, $z \equiv \tau r^{}_{1} + r^{}_{2}$ and $q^{}_z \equiv e^{2 {\rm i} \pi z}$. Now we take $k = 1$, then the basis vectors of the modular space ${\cal M}^{}_1 \left[\Gamma(5)\right]$ can be expressed as
\begin{eqnarray}
\begin{array}{lcl}
\widehat{e}^{}_{1}(\tau) = \dfrac{\eta^{15}(5 \tau)}{\eta^{3}(\tau)} \, \mathfrak{k}^{5}_{\frac{2}{5}, \frac{0}{5}}(5 \tau) \; , & ~\quad & \widehat{e}^{}_{2}(\tau) = \dfrac{\eta^{15}(5 \tau)}{\eta^{3}(\tau)} \, \mathfrak{k}^{}_{\frac{1}{5}, \frac{0}{5}}(5 \tau) \,  \mathfrak{k}^{4}_{\frac{2}{5}, \frac{0}{5}}(5 \tau) \; , \\
\widehat{e}^{}_{3}(\tau) = \dfrac{\eta^{15}(5 \tau)}{\eta^{3}(\tau)} \, \mathfrak{k}^{2}_{\frac{1}{5}, \frac{0}{5}}(5 \tau) \, \mathfrak{k}^{3}_{\frac{2}{5}, \frac{0}{5}}(5 \tau) \; , & ~\quad & \widehat{e}^{}_{4}(\tau) = \dfrac{\eta^{15}(5 \tau)}{\eta^{3}(\tau)} \, \mathfrak{k}^{3}_{\frac{1}{5}, \frac{0}{5}}(5 \tau) \, \mathfrak{k}^{2}_{\frac{2}{5}, \frac{0}{5}}(5 \tau) \; , \\
\widehat{e}^{}_{5}(\tau) = \dfrac{\eta^{15}(5 \tau)}{\eta^{3}(\tau)} \, \mathfrak{k}^{4}_{\frac{1}{5}, \frac{0}{5}}(5 \tau) \, \mathfrak{k}^{}_{\frac{2}{5}, \frac{0}{5}}(5 \tau) \; , & ~\quad & \widehat{e}^{}_{6}(\tau) = \dfrac{\eta^{15}(5 \tau)}{\eta^{3}(\tau)} \, \mathfrak{k}^{5}_{\frac{1}{5}, \frac{0}{5}}(5 \tau) \; .
\end{array}
\label{eq:basvec}
%     (A.4)
\end{eqnarray}
Furthermore, making use of Eqs.~(\ref{eq:Dedekindeta}) and (\ref{eq:Kexpansion}), we can derive the Fourier expansions of the above six basis vectors
\begin{eqnarray}
\widehat{e}^{}_1 & = & 1 + 3q + 4q^2_{} + 2q^3_{} + q^4_{} + 3q^5_{} + 6q^6_{} + 4q^7_{} - q^9_{} + \cdots \; , \nonumber \\
\widehat{e}^{}_2 & = & q^{1/5}_{} \left( 1 + 2q + 2q^2_{} + q^3_{} + 2q^4_{} + 2q^5_{} + 2q^6_{} + q^7_{} + 2q^8_{} + 2q^9_{} + \cdots \right) \nonumber \; , \\
\widehat{e}^{}_3 & = & q^{2/5}_{} \left( 1 + q + q^2_{} + q^3_{} + 2q^4_{} + q^6_{} + q^7_{} + 2q^8_{} + q^9_{} + \cdots \right) \; ,\nonumber \\
\widehat{e}^{}_4 & = & q^{3/5}_{} \left( 1 + q^2_{} + q^3_{} + q^4_{} - q^5_{} + 2q^6_{} + 2q^8_{} + q^9_{} + \cdots \right) \; ,  \nonumber \\
\widehat{e}^{}_5 & = & q^{4/5}_{} \left( 1 - q + 2q^2_{} + 2q^6_{} - 2q^7_{} + 2q^8_{} + q^9_{} +\cdots \right) \; , \nonumber \\
\widehat{e}^{}_6 & = & q \left( 1 - 2q + 4q^2_{} - 3q^3_{} + q^4_{} + 2q^5_{} - 2q^6_{} + 3q^8_{} - 2q^9_{} + \cdots \right) \; .
\label{eq:basisq}
%     (A.5)
\end{eqnarray}

\end{document}